\documentclass{article}

\usepackage[onecolumn]{emulateapj}

\newcommand{\bell}{\mbox{\boldmath$\ell$}}
\newcommand{\be}{\mbox{\boldmath$e$}}
\newcommand{\bff}{\mbox{\boldmath$f$}}
\newcommand{\br}{\mbox{\boldmath$r$}}
\newcommand{\bt}{\mbox{\boldmath$t$}}
\newcommand{\bG}{\mbox{\boldmath$G$}}
\newcommand{\bT}{\mbox{\boldmath$T$}}
\newcommand{\hatbe}{\mbox{\boldmath$\hat e$}}

\lefthead{Lubow \& Ogilvie}
\righthead{Tilting of protostellar disks}

\begin{document}
 
\submitted{To appear in the Astrophysical Journal}

\title{On the tilting of protostellar disks by resonant tidal effects}
\author{S. H. Lubow\altaffilmark{1}
 and G. I. Ogilvie\altaffilmark{1,2}}
\altaffiltext{1}{Space Telescope Science Institute,
  3700 San Martin Drive, Baltimore, MD 21218}
\altaffiltext{2}{Max-Planck-Institut f\"ur Astrophysik,
  Karl-Schwarzschild-Stra\ss e 1, Postfach 1523,
  D-85740 Garching bei M\"unchen, Germany}

\begin{abstract}
  We consider the dynamics of a protostellar disk surrounding a star
  in a circular-orbit binary system.  Our aim is to determine whether,
  if the disk is initially tilted with respect to the plane of the
  binary orbit, the inclination of the system will increase or
  decrease with time.  The problem is conveniently formulated in the
  binary frame in which the tidal potential of the companion star is
  static.  We may then consider a steady, flat disk that is aligned
  with the binary plane and investigate its linear stability with
  respect to tilting or warping perturbations.  The dynamics is
  controlled by the competing effects of the $m=0$ and $m=2$ azimuthal
  Fourier components of the tidal potential.  In the presence of
  dissipation, the $m=0$ component causes alignment of the system,
  while the $m=2$ component has the opposite tendency.  We find that
  disks that are sufficiently large, in particular those that extend
  to their tidal truncation radii, are generally stable and will
  therefore tend to alignment with the binary plane on a time-scale
  comparable to that found in previous studies.  However, the effect
  of the $m=2$ component is enhanced in the vicinity of resonances
  where the outer radius of the disk is such that the natural
  frequency of a global bending mode of the disk is equal to twice the
  binary orbital frequency.  Under such circumstances, the disk can be
  unstable to tilting and acquire a warped shape, even in the absence
  of dissipation.  The outer radius corresponding to the primary
  resonance is always smaller than the tidal truncation radius.  For
  disks smaller than the primary resonance, the $m=2$ component may be
  able to cause a very slow growth of inclination through the effect
  of a near resonance that occurs close to the disk center.  We
  discuss these results in the light of recent observations of
  protostellar disks in binary systems.
\end{abstract}

\keywords{accretion, accretion disks --- binaries: close ---
  hydrodynamics --- instabilities --- stars: pre-main sequence ---
  waves}

\section{Introduction}

The existence of disks around young stars was spectacularly confirmed
by direct images from the {\it Hubble Space Telescope\/} ({\it HST\/})
(McCaughrean \& O'Dell 1996; Burrows et~al. 1996).  Observations
suggest that young stars are usually found in binary systems and that
young binaries typically interact strongly with the disks that
surround the stars (Ghez, Neugebauer, \& Matthews 1993; Mathieu 1994;
Osterloh \& Beckwith 1995; Jensen, Mathieu, \& Fuller 1996).  There is
growing evidence that disks within a binary are sometimes inclined
with respect to the binary orbital plane.  Such a case may have been
seen in {\it HST\/} and Keck images of a disk in the young binary HK
Tau (Stapelfeldt et~al. 1998; Koresko 1998).

Suppose that a protostellar disk surrounds a star in a circular-orbit
binary system, and that the disk is tilted with respect to the binary
orbital plane.  The evolution of the disk is affected by the tidal
field of the companion star, as has been considered by Papaloizou \&
Terquem (1995).  Some features of their analysis were confirmed in
three-dimensional numerical simulations by Larwood et~al. (1996).  The
basic physics involved may be summarized as follows (see also Bate et
al. 2000).  In a non-rotating frame of reference centered on the star
about which the disk orbits, the companion star orbits at the binary
frequency $\Omega_{\rm b}$ and exerts a time-dependent tidal torque on
the disk.  This torque may be decomposed into a steady component and
an oscillatory component with a frequency of $2\Omega_{\rm b}$, and
their effects may be considered separately.

Consider first the steady torque.  If the disk were composed of
non-interacting circular rings, the steady torque would cause each
ring to precess, about an axis perpendicular to the binary plane, at a
rate that depends on the radius of the ring, resulting in a rapid
twisting of the disk.  However, if the disk is able to maintain
efficient radial communication, whether by wave propagation,
viscosity, or self-gravitation, it may be able to resist this
differential precession by establishing an internal torque in the
disk.  This can be arranged so that the net torque on each ring is
such as to produce a single, uniform precession rate.  However, to
establish this internal torque, the disk must become warped.  The
concomitant dissipation changes the total angular momentum of the
disk, tending to bring it into alignment with the binary plane in
addition to causing accretion.

Consider now the oscillatory torque.  Applied to a single ring, this
would cause a modulation of the precession rate and also a nutation
(Katz et~al. 1982).  However, in the presence of radial communication,
the oscillatory torque drives a bending wave (with azimuthal
wavenumber $m=1$) in the disk.  Papaloizou \& Terquem (1995) showed
that, if the wave is subject to dissipation, it too may change the
total angular momentum of the disk and tend to increase its
inclination.

The net effect of the steady and oscillatory torques determines
whether an initially coplanar disk will acquire a tilt over time or
whether an initially inclined disk will evolve towards coplanarity.
The purpose of this paper is to determine this outcome, which could
provide clues to the origin of misaligned disks in systems such as HK
Tau.

The basic mechanism suggested by Papaloizou \& Terquem for generating
a tilt by the oscillatory torque can be related to earlier work by
Lubow (1992), who showed that an aligned, Keplerian disk in a circular
binary may be linearly unstable to tilting if it contains a local
resonance at which the orbital angular velocity $\Omega(r)$ satisfies
\begin{equation}
  \Omega=\left({{m_*}\over{m_*-2}}\right)\Omega_{\rm b}.
  \label{resonance}
\end{equation}
Here $m_*$ is the azimuthal wavenumber of the component of the tidal
potential that is involved in the instability cycle.  The cycle works
through a mode-coupling process as follows: given a perturbation with
$m=1$ (a tilt), the tidal potential interacts with it to a drive a
wave with $m=m_*-1$ at the resonant radius.  This in turn interacts
with the tidal potential to produce a stress with $m=1$, which can
influence the tilt.  The role of dissipation is subtle, since some
dissipation is required to provide a change in the angular momentum of
the disk if instability is to occur, yet the associated damping can
compete with the intrinsic growth rate of the instability.

In particular, if the disk extends to the $3:1$ resonance
($\Omega=3\Omega_{\rm b}$) it may be unstable to tilting through the
$m=3$ component of the tidal potential.  This resonance has the
smallest $m_*$ for which equation (\ref{resonance}) can be satisfied
(for a prograde disk) and is the closest resonance to the central
star.  It is difficult for disks to extend even as far as the $3:1$
resonance, because of the effects of tidal truncation (Paczy\'nski
1977; Papaloizou \& Pringle 1977).  Superhump binary disks might
extend to the $3:1$ resonance because of their extreme binary mass
ratios, the secondary companion having less than $1/5$ the mass of the
primary about which the disk orbits (see the review by Osaki 1996).
This instability is related to, and occurs at the same position as,
the eccentric instability that is believed to be responsible for
superhumps in cataclysmic variable disks (Lubow 1991).  However, the
growth rate is invariably much smaller for tilting than for
eccentricity, and the weak tilt instability may be suppressed by the
effects of viscous damping and accretion (Murray \& Armitage 1998).
The same instabilities had been previously identified in the context
of planetary rings for higher $m_*$ (Goldreich \& Tremaine 1981;
Borderies, Goldreich, \& Tremaine 1984).  More fundamentally,
free-particle orbits undergo even stronger, parametric instabilities
at these resonant locations (Paczy\'nski 1977), although free
particles fail to model properly the behavior of a fluid disk at
resonances.

We relate this theory to the suggestion of Papaloizou \& Terquem
(1995) by noticing in equation (\ref{resonance}) that, for $m_*=2$, a
near resonance is obtained in the inner part of the disk where
$\Omega\gg\Omega_{\rm b}$.  Indeed, Papaloizou \& Terquem rely on the
$m=2$ component of the tidal potential to drive an $m=1$ bending wave
in the tilted disk.  The resulting response is a slowly rotating $m=1$
bending wave, with frequency $2\Omega_{\rm b}$ in the inertial frame.
Such a wave is close to resonance in the inner part of a nearly
Keplerian disk because of the near coincidence of the effective wave
driving frequency $\Omega-2\Omega_{\rm b}$ and the frequency of
vertical oscillations $\Omega_z\approx\Omega$; this is indeed the
origin of equation (\ref{resonance}) with $m_*=2$.  An additional
resonant effect occurs owing to the near coincidence of the driving
frequency and the epicyclic frequency of horizontal oscillations
$\kappa\approx\Omega$.  We describe the instability cycle associated
with the oscillatory torque as a mode-coupling process in Fig.~1.
However, because the resonance is not exact, and because of the
importance of resonantly induced horizontal motions, a proper
treatment requires a distinct analysis from that of Lubow (1992).

\placefigure{fig1}

In this paper, we therefore examine whether a flat, aligned disk in a
binary is linearly unstable to tilting even if it does not extend to
the $3:1$ resonance.  This problem is most conveniently analyzed in
the binary frame where the tidal potential is static, since the disk
can then be considered to be steady and to admit normal modes.  These
modes do not have a pure azimuthal wavenumber because the disk is
non-axisymmetric as a result of tidal distortions.  However, the
tilting instability, if present, may be expected to appear as a
modification of the rigid-tilt mode, which is trivial in the absence
of the companion star.  This mode may be followed continuously as the
mass of the companion is increased, in order to determine whether it
acquires a net rate of growth or decay.

In general, the analysis of a normal mode of a tidally distorted disk
is very difficult owing to the non-axisymmetric distortions of the
disk.  We therefore adopt the following simple approach, which is
appropriate when only $m=1$ bending waves are involved.  We start by
writing down the reduced equations for linear bending waves in a
protostellar disk subject to an axisymmetric external potential
(Section~2).  These can be derived formally without great effort (see
the Appendix).  We then give a physical interpretation of these
equations and use this insight to see how to modify them in the
presence of a non-axisymmetric potential (Sections~3 and~4).  We
present a simple disk model (Section~5) and describe the results of
numerical calculations of normal modes (Section~6).  Some further
analysis illuminates the underlying physics and helps to explain the
numerical results (Section~7).  Finally, we summarize our findings
(Section~8).

\section{Reduced description of linear bending waves}

Consider a thin, non-self-gravitating disk in an external
gravitational potential $\Phi(r,z)$ that is axisymmetric, but not
necessarily spherically symmetric.  Here $(r,\phi,z)$ are cylindrical
polar coordinates.  The orbital angular velocity $\Omega(r)$, the
epicyclic frequency $\kappa(r)$, and the vertical frequency
$\Omega_z(r)$ are defined by \footnote{The true angular velocity of
  the fluid will depart from $\Omega$ as a result of the radial
  pressure gradient and the vertical variation of the potential.  Such
  departures generally depend on $z$ and are of fractional order
  $(H/r)^2$.  They are fully taken into account in the analysis in the
  Appendix.}
\begin{eqnarray}
  \Omega^2&=&{{1}\over{r}}{{\partial\Phi}\over{\partial r}}\bigg|_{z=0},\\
  \kappa^2&=&4\Omega^2+2r\Omega{{d\Omega}\over{dr}},\\
  \Omega_z^2&=&{{\partial^2\Phi}\over{\partial z^2}}\bigg|_{z=0}.
\end{eqnarray}
We consider a situation in which the disk is nearly Keplerian and
almost inviscid in the sense that
\begin{eqnarray}
  \left|{{\kappa^2-\Omega^2}\over{\Omega^2}}\right|&\la&{{H}\over{r}},
  \label{criterion1}\\
  \left|{{\Omega_z^2-\Omega^2}\over{\Omega^2}}\right|&\la&{{H}\over{r}},\\
  \alpha&\la&{{H}\over{r}}\label{criterion3},
\end{eqnarray}
where $H(r)$ is the semi-thickness of the disk and $\alpha$ the
dimensionless viscosity parameter.  Then the linearized equations for
bending waves (with azimuthal wavenumber $m=1$) may be written
\notetoeditor{The symbol for the second vertical moment of the density
  ${\cal I}$ in this and subsequent equations should appear as a
  script I.}
\begin{equation}
  \Sigma r^2\Omega\left[{{\partial W}\over{\partial t}}+
  \left({{\Omega_z^2-\Omega^2}\over{\Omega^2}}\right)
  {{i\Omega}\over{2}}W\right]=
  {{1}\over{r}}{{\partial G}\over{\partial r}},
  \label{dwdt}
\end{equation}
\begin{equation}
  {{\partial G}\over{\partial t}}+
  \left({{\kappa^2-\Omega^2}\over{\Omega^2}}\right)
  {{i\Omega}\over{2}}G+\alpha\Omega G=
  {{{\cal I}r^3\Omega^3}\over{4}}{{\partial W}\over{\partial r}}.
  \label{dgdt}
\end{equation}
Here $\Sigma(r)$ is the surface density and ${\cal I}(r)$ the second
vertical moment of the density, defined by
\begin{equation}
  \Sigma=\int\rho\,dz,\qquad{\cal I}=\int\rho z^2\,dz.
\end{equation}
The second moment is related to the integrated pressure through
\begin{equation}
  \int p\,dz={\cal I}\Omega_z^2.
\end{equation}
The dimensionless complex variable $W(r,t)$ is defined by
$W=\ell_x+i\ell_y$, where $\bell(r,t)$ is the tilt vector, a unit
vector parallel to the local angular momentum vector of the disk.  The
complex variable $G(r,t)$ represents the internal torque which acts to
communicate stresses radially through the disk (see below).

The derivation of these equations may be found in the Appendix.
Equivalent equations, although presented in quite different notations,
have been derived by Papaloizou \& Lin (1995) and Demianski \& Ivanov
(1997).  In addition to conditions
(\ref{criterion1})--(\ref{criterion3}), it is required that the warp
vary on a length-scale long compared to the thickness of the disk, and
on a time-scale long compared to the local orbital time-scale.
However, any evolution of the disk on the (much longer) viscous
time-scale is neglected.  The (dynamic) viscosity is assumed to be
isotropic and proportional to the pressure ($\mu=\alpha p/\Omega$).

We emphasize the physical interpretation of these equations.  Equation
(\ref{dwdt}) contains the horizontal components of the angular
momentum equation encoded in the combination `$x+iy$'.  In vectorial
form it may be written
\begin{equation}
  \Sigma r^2\Omega{{\partial\bell}\over{\partial t}}=
  {{1}\over{r}}{{\partial\bG}\over{\partial r}}+\bT,
  \label{dldt}
\end{equation}
where $2\pi\bG(r,t)$ is the internal torque and $\bT(r,t)$ the
external torque density acting on the disk.  In the present case the
external torque arises from a lack of spherical symmetry in the
potential.  The complex variable $G$ is simply $G_x+iG_y$.  Its
equation may be written, in vectorial form,
\begin{equation}
  {{\partial\bG}\over{\partial t}}+
  \left({{\kappa^2-\Omega^2}\over{\Omega^2}}\right)
  {{\Omega}\over{2}}\be_z\times\bG+\alpha\Omega\bG=
  {{{\cal I}r^3\Omega^3}\over{4}}{{\partial\bell}\over{\partial r}}.
  \label{dgvecdt}
\end{equation}
The internal torque is mediated by horizontal epicyclic motions that
are driven near resonance by horizontal pressure gradients in the
warped disk, an effect identified by Papaloizou \& Pringle (1983).
The horizontal motions are proportional to $z$ and are therefore
subject to strong viscous dissipation which is the dominant channel of
damping of the bending waves.

We note that slowly varying $m=1$ bending waves or warps may be quite
generally described by conservation equations for mass and angular
momentum (Pringle 1992; Ogilvie 1999).  The relevant relation of $\bG$
to $\bell$ and its derivatives, however, depends strongly on the
thickness of the disk, the viscosity, and the rotation law.  In this
paper we are considering a parameter range appropriate to protostellar
disks, and are assuming that the warping is small so that a linear
theory is valid.  For disks in which $\alpha\ga(H/r)$, see Papaloizou
\& Pringle (1983), Pringle (1992), and Ogilvie (1999, 2000).

\section{Tidal torque on a tilted ring}

There are two dynamical degrees of freedom in the system described by
the above equations.  One is the tilting of the disk at each radius
according to the tilt vector $\bell$.  The other is the horizontal
motions described by $\bG$, which cause eccentric distortions of the
disk that are proportional to $z$.  In spite of this complexity, the
external torque density $\bT$ in equation (\ref{dldt}) is very simple
(cf. eq.~[\ref{dwdt}]): it is equal to the torque exerted by the
external potential on a disk composed of arbitrarily thin circular
rings of uniform density that are tilted according the tilt vector
$\bell$.  The eccentric distortions may be disregarded when
calculating the external torque to the required order.  \footnote{The
  effects of eccentric and tidal distortions are considered implicitly
  in Section 7, where they are found to be unimportant for the linear
  growth rates we derive.}

We proceed to derive an expression for the torque exerted by the full
potential of the companion star (of mass $M_2$) on a tilted ring of
the disk, treated as a thin and narrow circular ring of radius $r$ and
uniform density.  Adopt Cartesian coordinates $(x,y,z)$ with origin at
the center of the ring, and with the ring in the $xy$-plane.  Then the
position of an arbitrary point on the ring is
\begin{equation}
  \br=(r\cos\phi,r\sin\phi,0),
\end{equation}
where $\phi$ is the azimuthal angle measured around the ring.  Assume,
without loss of generality, that the companion star lies
instantaneously in the $xz$-plane at position
\begin{equation}
  \br_{\rm b}=(r_{\rm b}\cos\beta,0,r_{\rm b}\sin\beta),
\end{equation}
where $r_{\rm b}$ is the binary radius and $\beta$ the angle of
inclination.  Then the force per unit mass at position $\br$ on the
ring is
\begin{equation}
  \bff={{GM_2(\br_{\rm b}-\br)}\over{|\br_{\rm b}-\br|^3}},
\end{equation}
and the corresponding torque per unit mass is
\begin{equation}
  \bt=\br\times\bff=
  {{GM_2(\br\times\br_{\rm b})}\over{|\br_{\rm b}-\br|^3}}.
\end{equation}
The azimuthally averaged torque per unit mass is
\begin{equation}
  \langle\bt\rangle={{1}\over{2\pi}}\int_0^{2\pi}\bt\,d\phi.
\end{equation}
The $x$- and $z$-components vanish owing to the antisymmetry of the
integrands.  The remaining component is
\begin{equation}
  \langle t_y\rangle=
  -{{GM_2rr_{\rm b}\sin\beta}\over{2\pi}}\int_0^{2\pi}\cos\phi
  \left(r^2+r_{\rm b}^2-2rr_{\rm b}\cos\beta\cos\phi\right)^{-3/2}\,d\phi.
\end{equation}
In general, this may be expressed in terms of elliptic integrals.  For
small $\beta$, however, we have
\begin{equation}
  \langle t_y\rangle=-{{GM_2r\beta}\over{2r_{\rm b}^2}}
  \left[b_{3/2}^{(1)}\left({{r}\over{r_{\rm b}}}\right)\right]+
  O(\beta^3),
\end{equation}
where
\begin{equation}
  b_\gamma^{(m)}(x)={{2}\over{\pi}}\int_0^\pi\cos(m\phi)
  \left(1+x^2-2x\cos\phi\right)^{-\gamma}\,d\phi
\end{equation}
is the Laplace coefficient.  In vectorial form, therefore, the torque
density is
\begin{equation}
  \bT={{GM_2}\over{2r_{\rm b}^4}}
  \left[b_{3/2}^{(1)}\left({{r}\over{r_{\rm b}}}\right)\right]
  \Sigma r(\br_{\rm b}\cdot\bell)(\br_{\rm b}\times\bell),
\end{equation}
with fractional corrections of $O(W^2)$.

\section{Dynamics in the binary frame}

We now consider the dynamics in the binary frame, which rotates with
angular velocity ${\bf\Omega}_{\rm b}=\Omega_{\rm b}\,\be_z$.  This
requires that we replace, in equations (\ref{dldt}) and
(\ref{dgvecdt}),
\begin{equation}
  {{\partial\bell}\over{\partial t}}\mapsto
  {{\partial\bell}\over{\partial t}}+{\bf\Omega}_{\rm b}\times\bell,
  \qquad
  {{\partial\bG}\over{\partial t}}\mapsto
  {{\partial\bG}\over{\partial t}}+{\bf\Omega}_{\rm b}\times\bG.
\end{equation}
With the companion star located on the positive $x$-axis, we have, in
linear theory,
\begin{equation}
  \bT=-{{GM_2}\over{2r_{\rm b}^2}}
  \left[b_{3/2}^{(1)}\left({{r}\over{r_{\rm b}}}\right)\right]
  \Sigma r\ell_x\,\be_y.
\end{equation}
The orbital, epicyclic, and vertical frequencies are all calculated
using the $m=0$ total potential.  This gives
\begin{eqnarray}
  \Omega^2&=&{{GM_1}\over{r^3}}+
  {{GM_2}\over{2r_{\rm b}^2r}}\left[{{r}\over{r_{\rm b}}}
  b_{3/2}^{(0)}\left({{r}\over{r_{\rm b}}}\right)-
  b_{3/2}^{(1)}\left({{r}\over{r_{\rm b}}}\right)\right],\\
  \kappa^2&=&{{GM_1}\over{r^3}}+
  {{GM_2}\over{2r_{\rm b}^2r}}\left[{{r}\over{r_{\rm b}}}
  b_{3/2}^{(0)}\left({{r}\over{r_{\rm b}}}\right)-
  2b_{3/2}^{(1)}\left({{r}\over{r_{\rm b}}}\right)\right],\\
  \Omega_z^2&=&{{GM_1}\over{r^3}}+
  {{GM_2}\over{2r_{\rm b}^2r}}\left[{{r}\over{r_{\rm b}}}
  b_{3/2}^{(0)}\left({{r}\over{r_{\rm b}}}\right)\right],
\end{eqnarray}
where $M_1$ is the mass of the star about which the disk orbits.  We
assume, without loss of generality, that $\Omega>0$, but allow for the
orbit of the companion star to be either prograde or retrograde
according to
\begin{equation}
  \Omega_{\rm b}=\pm\left[{{G(M_1+M_2)}\over{r_{\rm b}^3}}\right]^{1/2}.
\end{equation}

The final equations are
\begin{equation}
  \Sigma r^2\Omega\left({{\partial\ell_x}\over{\partial t}}-
  \Omega_{\rm b}\ell_y\right)=
  {{1}\over{r}}{{\partial G_x}\over{\partial r}},
  \label{ell_x}
\end{equation}
\begin{equation}
  \Sigma r^2\Omega\left({{\partial\ell_y}\over{\partial t}}+
  \Omega_{\rm b}\ell_x\right)=
  {{1}\over{r}}{{\partial G_y}\over{\partial r}}-
  {{GM_2}\over{2r_{\rm b}^2}}
  \left[b_{3/2}^{(1)}\left({{r}\over{r_{\rm b}}}\right)\right]
  \Sigma r\ell_x,
\end{equation}
\begin{equation}
  {{\partial G_x}\over{\partial t}}-\Omega_{\rm b}G_y+
  {{GM_2}\over{4r_{\rm b}^2 r\Omega}}
  \left[b_{3/2}^{(1)}\left({{r}\over{r_{\rm b}}}\right)\right]G_y+
  \alpha\Omega G_x=
  {{{\cal I}r^3\Omega^3}\over{4}}{{\partial\ell_x}\over{\partial r}},
\end{equation}
\begin{equation}
  {{\partial G_y}\over{\partial t}}+\Omega_{\rm b}G_x-
  {{GM_2}\over{4r_{\rm b}^2 r\Omega}}
  \left[b_{3/2}^{(1)}\left({{r}\over{r_{\rm b}}}\right)\right]G_x+
  \alpha\Omega G_y=
  {{{\cal I}r^3\Omega^3}\over{4}}{{\partial\ell_y}\over{\partial r}}.
  \label{G_y}
\end{equation}

Since the coefficients of these equations are independent of time, we
may seek normal modes of the form
\begin{eqnarray}
  \ell_x(r,t)&=&{\rm Re}\left[\tilde\ell_x(r)\,e^{i\omega t}\right],
  \label{tilde_ell_x}\\
  \ell_y(r,t)&=&{\rm Re}\left[\tilde\ell_y(r)\,e^{i\omega t}\right],\\
  G_x(r,t)&=&{\rm Re}\left[\tilde G_x(r)\,e^{i\omega t}\right],\\
  G_y(r,t)&=&{\rm Re}\left[\tilde G_y(r)\,e^{i\omega t}\right],
  \label{tilde_G_y}
\end{eqnarray}
where $\omega$ is a complex frequency eigenvalue.  The problem has
then been reduced to solving an eigenvalue problem involving a
fourth-order system of ordinary differential equations (ODEs).

If we return to the original complex notation, we find that the
equations have become non-analytic, effectively increasing the order
of the dynamical system:
\begin{equation}
  \Sigma r^2\Omega\left({{\partial W}\over{\partial t}}+
  i\Omega_{\rm b}W\right)=
  {{1}\over{r}}{{\partial G}\over{\partial r}}-
  {{GM_2}\over{4r_{\rm b}^2}}
  \left[b_{3/2}^{(1)}\left({{r}\over{r_{\rm b}}}\right)\right]
  \Sigma ri(W+W^*),
\end{equation}
\begin{equation}
  {{\partial G}\over{\partial t}}+i\Omega_{\rm b}G-
  {{GM_2}\over{4r_{\rm b}^2 r\Omega}}
  \left[b_{3/2}^{(1)}\left({{r}\over{r_{\rm b}}}\right)\right]iG+
  \alpha\Omega G=
  {{{\cal I}r^3\Omega^3}\over{4}}{{\partial W}\over{\partial r}}.
\end{equation}
In the combination $W+W^*$, the term $W$ arises from the $m=0$
component of the tidal potential (cf. eq.~[\ref{dwdt}]), while the
non-analytic term $W^*$ arises from the $m=2$ component.  In the
normal-mode solution, $W$ and $G$ have the form
\begin{eqnarray}
  W&=&W_+\,e^{i\omega t}+W_-\,e^{-i\omega^*t},
  \label{wpm}\\
  G&=&G_+\,e^{i\omega t}+G_-\,e^{-i\omega^*t},
  \label{gpm}
\end{eqnarray}
where
\begin{eqnarray}
  W_+&=&{\textstyle{{1}\over{2}}}(\tilde\ell_x+i\tilde\ell_y),\\
  W_-&=&{\textstyle{{1}\over{2}}}(\tilde\ell_x^*+i\tilde\ell_y^*),\\
  G_+&=&{\textstyle{{1}\over{2}}}(\tilde G_x+i\tilde G_y),\\
  G_-&=&{\textstyle{{1}\over{2}}}(\tilde G_x^*+i\tilde G_y^*).
\end{eqnarray}

The motion seen in the inertial frame is more complicated than a
single mode.  We have
\begin{equation}
  \bell=\ell_x\,\be_x+\ell_y\,\be_y+\ell_z\,\be_z,
\end{equation}
where $(\be_x,\be_y,\be_z)$ are unit vectors in the binary frame.
These are related to the unit vectors $(\hatbe_x,\hatbe_y,\be_z)$ in
the inertial frame by
\begin{equation}
  \be_x-i\be_y=(\hatbe_x-i\hatbe_y)\,e^{i\Omega_{\rm b}t},
\end{equation}
and so
\begin{equation}
  \bell=\hat\ell_x\,\hatbe_x+\hat\ell_y\,\hatbe_y+\ell_z\,\be_z,
\end{equation}
where
\begin{equation}
  \hat\ell_x+i\hat\ell_y=W\,e^{i\Omega_{\rm b}t}=
  e^{-\omega_{\rm i}t}
  \left[W_+\,e^{i(\omega_{\rm r}+\Omega_{\rm b})t}+
  W_-\,e^{-i(\omega_{\rm r}-\Omega_{\rm b})t}\right].
  \label{hatl}
\end{equation}
Here $\omega=\omega_{\rm r}+i\omega_{\rm i}$.  Therefore two
components are seen in the inertial frame, which have distinct
frequencies, $|\omega_{\rm r}\pm\Omega_{\rm b}|$, but the same rate of
growth or decay.

\section{Disk model}

For simplicity, we assume that the vertical structure of the
unperturbed disk is that of a polytrope of index $n$.  To satisfy
vertical hydrostatic equilibrium, the density distribution, for a thin
disk, is then of the form
\begin{equation}
  \rho(r,z)=\rho(r,0)\left(1-{{z^2}\over{H^2}}\right)^n,
\end{equation}
where $H(r)$ is the semi-thickness.  The surface density and second
moment are related by
\begin{equation}
  {\cal I}={{\Sigma H^2}\over{2n+3}}.  
\end{equation}
For the radial structure, we specify
\begin{equation}
  {{H}\over{r}}=\epsilon
\end{equation}
and
\begin{equation}
  \Sigma=\Sigma_0r^{-1/2}f,
\end{equation}
where $\epsilon$ is a small constant, $\Sigma_0$ an arbitrary
constant, and $f(r)$ a function that is approximately equal to unity
except near the inner and outer radii of the disk, where it tapers
linearly to zero.  Over most of the disk this gives approximately
$H\propto r$, $\Sigma\propto r^{-1/2}$, and ${\cal I}\propto r^{3/2}$.

For the tapering function, we take
\begin{equation}
  f=\tanh\left({{r-r_1}\over{w_1}}\right)
  \tanh\left({{r_2-r}\over{w_2}}\right),
\end{equation}
where $r_1$ and $r_2$ are the inner and outer radii of the disk, and
$w_1$ and $w_2$ are the widths of the tapers near each edge, which are
taken to be equal to the local semi-thickness.  With $f$ tapering
linearly to zero, the edges are regular singular points of the
governing equations.  The appropriate boundary condition in each case
is that $W$ should be regular there, which implies that $G$ vanishes.
Clearly the internal torque cannot be transmitted across a free
boundary of the disk.  However, if the inner disk were terminated by a
magnetosphere, for example, this boundary condition may require
modification.

This model is very similar to that used by Papaloizou \& Terquem
(1995) except that the disk has an inner edge.  For reasons that we
explain in Section~7, we do not attempt to impose an `ingoing wave'
boundary condition at the center of the disk.

\begin{deluxetable}{lcr}
  \tablecaption{Parameters of the reference model.
  \label{tab1}}
  \tablewidth{0pt}
  \tablehead{\colhead{Parameter}&\colhead{Symbol}&\colhead{Value}}
  \startdata
    Mass ratio&$q=M_2/M_1$&$1$\nl
    Angular semi-thickness&$\epsilon$&$0.1$\nl
    Viscosity parameter&$\alpha$&$0.01$\nl
    Inner radius&$r_1/r_{\rm b}$&$0.01$\nl
    Outer radius&$r_2/r_{\rm b}$&$0.3$\nl
    Width of inner taper&$w_1/r_1$&$0.1$\nl
    Width of outer taper&$w_2/r_2$&$0.1$\nl
    Polytropic index&$n$&$3/2$\nl
  \enddata
\end{deluxetable}

\begin{deluxetable}{lrlr}
  \tablecaption{Frequency eigenvalues for the reference model, but with
  $q=0$ and $\alpha=0$.
  \label{tab2}}
  \tablewidth{0pt}
  \tablehead{\colhead{Mode}&\colhead{$\omega/\Omega_{\rm b}$}&
  \colhead{Mode}&\colhead{$\omega/\Omega_{\rm b}$}}
  \startdata
    $0$&$-1$&$0^*$&$1$\nl
    $1_+$&$-0.3074$&$1_+^*$&$+0.3074$\nl
    $1_-$&$-1.6926$&$1_-^*$&$+1.6926$\nl
    $2_+$&$+0.3163$&$2_+^*$&$-0.3163$\nl
    $2_-$&$-2.3163$&$2_-^*$&$+2.3163$\nl
    $3_+$&$+0.9213$&$3_+^*$&$-0.9213$\nl
    $3_-$&$-2.9213$&$3_-^*$&$+2.9213$\nl
    $4_+$&$+1.5180$&$4_+^*$&$-1.5180$\nl
    $4_-$&$-3.5180$&$4_-^*$&$+3.5180$\nl
  \enddata
\end{deluxetable}

\begin{deluxetable}{lr}
  \tablecaption{Frequency eigenvalues for the reference model, but
    with $q=0$ and $\alpha = 0.01$.
  \label{tab3}}
  \tablewidth{0pt}
  \tablehead{\colhead{Mode}&\colhead{$\omega/\Omega_{\rm b}$}}
  \startdata
    $0$&$-1$\nl
    $1_+$&$-0.3101+0.0752i$\nl
    $1_-$&$-1.6899+0.0752i$\nl
    $2_+$&$+0.3144+0.0925i$\nl
    $2_-$&$-2.3144+0.0925i$\nl
    $3_+$&$+0.9197+0.1020i$\nl
    $3_-$&$-2.9197+0.1020i$\nl
    $4_+$&$+1.5166+0.1087i$\nl
    $4_-$&$-3.5166+0.1087i$\nl
  \enddata
\end{deluxetable}

\section{Numerical results}

Equations (\ref{ell_x})--(\ref{G_y}) are solved numerically using the
complex variables defined in equations
(\ref{tilde_ell_x})--(\ref{tilde_G_y}).  When solving the ODEs for a
normal mode, it is advisable to integrate away from the singular
points at the edges of the disk.  We apply the arbitrary normalization
condition $\tilde\ell_x(r_1)=1$ and guess the values of the four
complex parameters $\omega$, $\tilde\ell_y(r_1)$, $\tilde\ell_x(r_2)$,
and $\tilde\ell_y(r_2)$.  We then integrate separately into $r>r_1$
and $r<r_2$, meeting at the midpoint, where $\tilde\ell_x$,
$\tilde\ell_y$, $\tilde G_x$, and $\tilde G_y$ should all be
continuous.  These four conditions are solved by Newton-Raphson
iteration, using derivative information obtained by simultaneously
integrating the ODEs differentiated with respect to the four
parameters.

\subsection{Reference model}

We first identify a `reference model' with parameters that we consider
appropriate for a protostellar disk that is tidally truncated by the
companion star (Table~1).  The orbit of the companion is taken to be
prograde.

Before considering the reference model as such, we examine the same
disk but with no viscosity ($\alpha=0$) and with a companion of zero
mass ($q=0$).  An infinite set of discrete bending modes is obtained,
which are characterized by the number of nodes in the eigenfunction
$\tilde\ell_x$ (say).  The basic frequencies of these modes in the
inertial frame are $\omega_0=0$, $\omega_1=0.6926\,\Omega_{\rm b}$,
$\omega_2=1.3163\,\Omega_{\rm b}$, $\omega_3=1.9213\,\Omega_{\rm b}$,
$\omega_4=2.5180\,\Omega_{\rm b}$, etc.  We refer to these modes as
modes $0$, $1$, $2$, $3$, $4$, etc.  Mode $0$ is the (trivial)
rigid-tilt mode and has no nodes.

In the binary frame, the full set of frequencies appears much more
complicated, as shown in Table~2.  The modes in the left-hand column
consist purely of $W_+$ and $G_+$, having $W_-=0$ and $G_-=0$.  For
such a mode, the frequency in the binary frame is less than the
frequency in the inertial frame by an amount $\Omega_{\rm b}$ (cf.
eq.~[\ref{hatl}]).  Since, in the inertial frame, we may have a
prograde or retrograde mode $n$ with frequency $\pm\omega_n$, we
obtain frequencies $\pm\omega_n-\Omega_{\rm b}$ in the binary frame.
These are labeled $n_\pm$.  The modes in the right-hand column are
physically equivalent.  The eigenfunctions and eigenvalues are
obtained from those in the left-hand column by complex conjugation and
a change of sign.  Such modes consist purely of $W_-$ and $G_-$,
having $W_+=0$ and $G_+=0$.  Thus the frequencies in the binary frame
are $\mp\omega_n+\Omega_{\rm b}$.  These modes are labeled $n_\pm^*$.

We consider next the effect of a small viscosity on the modes by
increasing $\alpha$ from $0$ to its reference value $0.01$, but still
with a companion of zero mass ($q=0$).  The results are shown in
Table~3.  We omit the complex-conjugate modes from now on, but their
existence should not be forgotten.  Evidently the real part of the
frequency changes very little in the presence of a small viscosity,
but, with the exception of the rigid-tilt mode, the frequency acquires
a positive imaginary part, which signifies a damping rate.  The
damping rate depends relatively little on the order of the mode.  It
can be seen from the governing equations that the effect of viscosity
is simply to damp the horizontal motions locally at a rate
$\alpha\Omega$ (cf. eq.~[\ref{dgvecdt}]).  Since the horizontal
motions are an essential part of each proper bending mode, this leads
to a damping rate for each mode of order $\alpha\Omega$ (evaluated in
the outer parts of the disk).  The exception is the rigid-tilt mode,
for which the horizontal motions are exactly zero.

Finally, we reach the reference model by increasing the binary mass
ratio $q$ from $0$ to its reference value $1$.  We start with mode
$0$, which corresponds to a rigid tilt and consists purely of $W_+$.
The frequency of the mode (now the `modified' rigid-tilt mode) changes
continuously from $-\Omega_{\rm b}$ to $(-1.0484+0.000258i)\Omega_{\rm
  b}$.  The mode also acquires a $W_-$ component.  Viewed in the
inertial frame, the mode changes from a pure $W_+$ mode with zero
frequency to a combination of $W_+$ and $W_-$ contributions having
frequencies of $0.0484\,\Omega_{\rm b}$ and $2.0484\,\Omega_{\rm b}$,
respectively (see eq.~[\ref{hatl}]).  The first frequency corresponds
to a retrograde precession of the tilted disk, forced by the $m=0$
component of the tidal potential.  The second corresponds to the
forcing of a bending wave ($W_-$) by the $m=2$ component of the
potential.  The two potential components provide the `steady' and
`oscillatory' torques, respectively.  Since the imaginary part of the
frequency is positive, the whole pattern decays at a rate
$0.000258\,\Omega_{\rm b}$.  The other modes of the disk are of course
damped much more rapidly, and we conclude that the reference model
disk is linearly stable to tilting.

\subsection{Resonances}

We now search the parameter space around the reference model for any
regions of instability.  In particular, we try varying the outer
radius $r_2$ of the disk.  In Fig.~2 we plot the dimensionless growth
rate $-\omega_{\rm i}/\Omega_{\rm b}$ against $r_2/r_{\rm b}$ for a
number of different values of $\alpha$.  It is clear that the net
growth rate is a combination of two parts.  One part is a damping
($\omega_{\rm i}>0$) that is proportional to $\alpha$ and increases
rapidly with increasing $r_2$.  The second part is a growth
($\omega_{\rm i}<0$) with an entirely different behavior.  The growth
is localized in a sequence of peaks which become higher and narrower
as $\alpha$ decreases.  In Fig.~3 we show an expanded view of the
primary peak for the cases $\alpha=0$ and $\alpha=0.001$.

\placefigure{fig2}

\placefigure{fig3}

To verify the origin of the two parts, we repeated the calculation
using equations that retain only the $m=0$ component of the tidal
potential, or only the $m=2$ component.  It is obvious from this that
the damping is due entirely to the $m=0$ component, while the growth
is due entirely to the $m=2$ component.  There is a slight shift in
the positions of the peaks when the $m=0$ component of the tidal
potential is neglected.

It is evident that the growth (that is, the instability) is associated
with a series of resonances that occur when the outer radius of the
disk is in the vicinity of certain discrete values.  In the absence of
viscosity, the resonances come about as follows.  As $r_2/r_{\rm b}$
is varied, the frequency eigenvalues of all bending modes migrate
along the real axis in the $\omega$-plane.  With the exception of mode
$0$, all modes are very sensitive to the position of the outer
boundary, which reflects the waves.  As a result, collisions occur on
the real axis.  In particular, when $r_2/r_{\rm b}$ is increased from
$0.1$ towards the primary resonance, mode $0$ undergoes a collision
with mode $1_+^*$ (a bending mode with one node).  The modes move
briefly off the real axis, producing a complex-conjugate pair, and
then return to the real axis to continue their original migration.
The other resonances occur when mode $0$ undergoes collisions with
modes $2_+^*$, $3_+^*$, etc.  During a collision, the two modes
exchange characteristics, and the eigenfunctions are hybrids of the
two original ones.  In particular, mode $0$ no longer resembles a
rigid tilt during a collision with a proper bending mode.  This means
that a disk made unstable by this means would develop a warped shape
(see Section 6.4 below).

In the presence of a very small viscosity, the proper bending modes
are damped and their eigenvalues are displaced somewhat above the real
axis.  The collisions are no longer exact and each mode can be
followed continuously as $r_2/r_{\rm b}$ is varied.  For
$\alpha=0.001$, say, the modes pass sufficiently close that a strong
interaction occurs.  The tracks of the eigenvalues are deflected to
avoid a collision, and, in so doing, mode $0$ acquires a positive
growth rate that appears as a resonance.  During the interaction, the
eigenfunction of mode $0$ is distorted significantly from a rigid
tilt, but not so strongly as in the inviscid case (see Section~6.4
below).

When the viscosity is increased, the resonances become broader and
weaker.  A positive growth rate is not achieved if the height of the
resonance is less than the damping rate arising from the $m=0$
potential.  Therefore the regions of instability are suppressed as
$\alpha$ is increased.  It appears that, as long as the primary
resonance survives, the net growth rate (for $\alpha>0$) is always
positive for disks smaller than the size of the primary resonance,
although the growth rate may be minuscule.  This may be considered as
a long tail of the primary resonance.  However, the primary peak is
dramatically reduced in height as $\alpha$ is increased, and it also
shifts to smaller radius.  In the cases investigated here, all traces
of instability are eliminated when $\alpha=0.1$.

To elucidate further the condition for resonance, we examined the
bending modes at their points of collision with mode $0$ and evaluated
their natural frequencies (i.e. in the absence of the tidal potential,
and evaluated in the inertial frame).  In each case the natural
frequency is close to $2\Omega_{\rm b}$ at the point of collision.
The resonances therefore occur when the oscillatory torque due to the
$m=2$ potential resonates with a free bending mode of the disk.

We remark that the global resonant excitation of bending waves has
been identified by Terquem (1998) when calculating the tidal torque
exerted on a protostellar disk by a companion in an inclined circular
orbit.  However, the consequences for the evolution of the relative
inclination of the system were not investigated.

The results for a companion in a retrograde orbit are not
significantly different.  The heights of the resonant peaks are very
similar, but they are shifted slightly in radius.  The shift of the
resonances (also observed, as noted above, when the $m=0$ component of
the tidal potential is omitted) is related to the precession of the
disk, which changes the effective frequency of the oscillatory torque
and, therefore, the condition for resonance.  The precession is always
retrograde in the inertial frame, irrespective of the sense of the
companion's orbit.  Therefore the effective driving frequency depends
on the sense of the orbit, but the shift is generally small.

\subsection{Precession rate and decay rate}

In Fig.~4 we plot the precession rate of the modified rigid-tilt mode
against the outer radius of the disk, for the reference model.  The
precession is always retrograde and the rate increases rapidly with
increasing $r_2$.  Excellent agreement is found with the simple
analytic approximation given by Bate et~al. (2000; eq.~[22]).  (We
have set the dimensionless parameter $K=0.4$, since this represents
fairly accurately the disk models we have adopted.)  For much smaller
values of $\alpha$, a noticeable deviation from this curve occurs in
the vicinity of resonances, since the path of the eigenvalue in the
$\omega$-plane is temporarily diverted.

\placefigure{fig4}

In Fig.~5 we plot the decay rate of the modified rigid-tilt mode, for
the reference model.  When only the $m=0$ component of the tidal
potential is included, the decay rate is always positive and increases
rapidly with increasing $r_2$.  When the full potential is used, the
behavior is modified in the vicinity of resonances.  Also shown is the
simple estimate $1/t_{\rm align}$ given by Bate et~al. (2000;
eq.~[35]).  Apart from the resonances, the simple estimate captures
the correct dependence on $r_2$.  It should be borne in mind that the
estimate of Bate et~al. (2000) was based on an order-of-magnitude
analysis, and can be expected to be accurate only within a factor of
order unity.

\placefigure{fig5}

\subsection{Shape of the disk}

For comparison with observations, it is of interest to examine the
shape adopted by the disk while executing the modified rigid-tilt
mode.  Information on the shape of the disk is contained in four real
functions of radius, namely the real and imaginary parts of the
eigenfunctions $\tilde\ell_x(r)$ and $\tilde\ell_y(r)$.  We display
this information in Figs~6 and~7 by showing cross-sections through the
disk in the $xz$- and $yz$-planes at two instants, corresponding to
phase $0$ and phase $\pi/2$ of the period seen in the binary frame.

Fig.~6 is for a disk with $r_2/r_{\rm b}=0.118$, in the middle of the
primary resonance.  Three different viscosities, $\alpha=0$,
$\alpha=0.001$, and $\alpha=0.01$, are considered.  In each case the
mode has a positive growth rate.  In the absence of viscosity, the
resonance is strong and the disk becomes distinctly warped in a smooth
and global manner.  As already noted, when viscosity is included, the
resonance is much weaker and the disk appears tilted with less
noticeable warping.

\placefigure{fig6}

Fig.~7 is for a disk with the reference value $r_2/r_{\rm b}=0.3$
representative of a tidally truncated disk.  We fix $\alpha=0.01$ and
consider disks of varying thickness, $\epsilon=0.1$, $\epsilon=0.05$,
and $\epsilon=0.03$.  In each case the mode is damped.  For
$\epsilon=0.1$, the disk appears tilted without noticeable warping.
For thinner disks, the deviation from a rigid tilt is noticeable in
the outer part of the disk where the tidal forcing is strongest.

\placefigure{fig7}

Recall that the derivation of equations (\ref{dldt}) and
(\ref{dgvecdt}) requires that the warp vary on a length-scale long
compared to the thickness of the disk (see the Appendix).  This
condition is indeed satisfied in the solutions we present here.

\section{Expansion in the tidal potential}

\subsection{Basic equations}

The normal-mode description affords an especially compact
representation of the dynamics and is very suitable for the numerical
analysis.  In this section we `unpack' the eigenfunction to reveal the
essential physics of the problem.  We write the basic equations in the
general form
\begin{equation}
  \Sigma r^2\Omega\left({{\partial W}\over{\partial t}}+
  i\Omega_{\rm b}W\right)=
  {{1}\over{r}}{{\partial G}\over{\partial r}}-i(AW+BW^*),
\end{equation}
\begin{equation}
  {{\partial G}\over{\partial t}}+i\Omega_{\rm b}G-i(CG+DG^*)+
  \alpha\Omega G=
  {{{\cal I}r^3\Omega^3}\over{4}}{{\partial W}\over{\partial r}},
\end{equation}
with unspecified coefficients $A$, $B$, $C$, and $D$ arising from the
tidal potential.  In view of our earlier discussion, terms $A$ and $C$
are due to the $m=0$ component of the potential, while the
non-analytic terms $B$ and $D$ are due to the $m=2$ component.  We
allow for the possibility that tidal distortions of the disk may
introduce additional complexities (such as a term $D$) that we have
not foreseen.

For a normal mode of the form (\ref{wpm})--(\ref{gpm}), we have
\begin{eqnarray}
  (i\omega+i\Omega_{\rm b})\Sigma r^2\Omega W_+&=&
  {{1}\over{r}}{{dG_+}\over{dr}}-i(AW_++BW_-^*),\\
  (-i\omega^*+i\Omega_{\rm b})\Sigma r^2\Omega W_-&=&
  {{1}\over{r}}{{dG_-}\over{dr}}-i(AW_-+BW_+^*),
\end{eqnarray}
\begin{eqnarray}
  (i\omega+i\Omega_{\rm b})G_+-i(CG_++DG_-^*)+\alpha\Omega G_+&=&
  {{{\cal I}r^3\Omega^3}\over{4}}{{dW_+}\over{dr}},\\
  (-i\omega^*+i\Omega_{\rm b})G_--i(CG_-+DG_+^*)+\alpha\Omega G_-&=&
  {{{\cal I}r^3\Omega^3}\over{4}}{{dW_-}\over{dr}}.
\end{eqnarray}

\subsection{Expansions}

We now expand the equations in powers of the tidal potential,
indicated by a numerical superscript.  The unspecified coefficients
may be assumed to have expansions
\begin{equation}
  A=A^{(1)}+A^{(2)}+\cdots,
\end{equation}
etc., since they vanish in the absence of the tidal potential.  The
eigenvalue and eigenfunction have the expansions
\begin{eqnarray}
  \omega&=&\omega^{(0)}+\omega^{(1)}+\omega^{(2)}+\cdots,\\
  W_+&=&W_+^{(0)}+W_+^{(1)}+W_+^{(2)}+\cdots,\\
  W_-&=&\phantom{W_+^{(0)}+{}}W_-^{(1)}+W_-^{(2)}+\cdots,\\
  G_+&=&\phantom{W_+^{(0)}+{}}G_+^{(1)}+G_+^{(2)}+\cdots,\\
  G_-&=&\phantom{W_+^{(0)}+{}}G_-^{(1)}+G_-^{(2)}+\cdots,
\end{eqnarray}
where, at leading order, we have the rigid-tilt mode with
\begin{equation}
  \omega^{(0)}=-\Omega_{\rm b},\qquad W_+^{(0)}={\rm constant}.
\end{equation}
The rigid-tilt amplitude could be arbitrarily specified as
$W_+^{(0)}=1$, but we retain $W_+^{(0)}$ for clarity in the equations
below.

\subsection{Solution}

At first order, we obtain
\begin{equation}
  i\omega^{(1)}\Sigma r^2\Omega W_+^{(0)}=
  {{1}\over{r}}{{dG_+^{(1)}}\over{dr}}-iA^{(1)} W_+^{(0)},
  \label{expand1}
\end{equation}
\begin{equation}
  2i\Omega_{\rm b}\Sigma r^2\Omega W_-^{(1)}=
  {{1}\over{r}}{{dG_-^{(1)}}\over{dr}}-iB^{(1)}W_+^{(0)*},
  \label{expand2}
\end{equation}
\begin{equation}
  \alpha\Omega G_+^{(1)}=
  {{{\cal I}r^3\Omega^3}\over{4}}{{dW_+^{(1)}}\over{dr}},
  \label{expand3}
\end{equation}
\begin{equation}
  (2i\Omega_{\rm b}+\alpha\Omega)G_-^{(1)}=
  {{{\cal I}r^3\Omega^3}\over{4}}{{dW_-^{(1)}}\over{dr}}.
  \label{expand4}
\end{equation}
From equation (\ref{expand1}), using the fact that $G$ vanishes at the
edges of the disk, we immediately obtain the solvability condition
\begin{equation}
  \omega^{(1)}=-\int_{r_1}^{r_2}A^{(1)}W_+^{(0)}\,r\,dr\bigg/
  \int_{r_1}^{r_2}\Sigma r^2\Omega W_+^{(0)}\,r\,dr,
\end{equation}
which relates the precession rate (at first order) to the total
horizontal tidal torque on the disk (at first order) divided by the
horizontal angular momentum of the tilted disk.  This effect is due to
$A$ and therefore to the $m=0$ component of the tidal potential.
Equations (\ref{expand1})--(\ref{expand4}) can then, in principle, be
solved for $W_\pm^{(1)}$ and $G_\pm^{(1)}$.

At second order, we obtain
\begin{equation}
  i\omega^{(2)}\Sigma r^2\Omega W_+^{(0)}+
  i\omega^{(1)}\Sigma r^2\Omega W_+^{(1)}=
  {{1}\over{r}}{{dG_+^{(2)}}\over{dr}}-
  i\left(A^{(2)}W_+^{(0)}+A^{(1)}W_+^{(1)}+B^{(1)}W_-^{(1)*}\right),
\end{equation}
plus three further equations, which will not be required.  This time
the solvability condition is
\begin{equation}
  \omega^{(2)}=
  -\int_{r_1}^{r_2}\left(A^{(2)}W_+^{(0)}+A^{(1)}W_+^{(1)}+
  \omega^{(1)}\Sigma r^2\Omega W_+^{(1)}+B^{(1)}W_-^{(1)*}\right)
  \,r\,dr\bigg/
  \int_{r_1}^{r_2}\Sigma r^2\Omega W_+^{(0)}\,r\,dr
  \label{omega2}. 
\end{equation}
If, as we assume, $A$ is real, then $\omega^{(1)}$ is real.  After
some further manipulations we then obtain
\begin{equation}
  {\rm Im}\left(\omega^{(2)}\right)=
  \int_{r_1}^{r_2}\left({{4\alpha}\over{{\cal I}r^4\Omega^2}}\right)
  \left(\left|G_+^{(1)}\right|^2-\left|G_-^{(1)}\right|^2\right)
  \,r\,dr\bigg/
  \int_{r_1}^{r_2}\Sigma r^2\Omega\left|W_+^{(0)}\right|^2\,r\,dr.
\end{equation}
This shows that $G_+^{(1)}$, which is caused by the $m=0$ component of
the potential, causes pure damping, while $G_-^{(1)}$, which is caused
by the $m=2$ component, causes pure growth.  The net effect depends on
which is larger in the norm defined above.  Note that coefficients $C$
and $D$ have no effect to this order.  Also, the second-order
coefficient $A^{(2)}$, which we did not attempt to calculate, does not
affect the growth or decay rate at second order (although it does
affect the precession frequency at second order).  Such a coefficient
could arise because the tidal torque on a tilted ring may have a
second-order correction owing to the tidal distortion of the ring.

Furthermore, since $G_+^{(1)}$ is independent of $\alpha$ according to
equation (\ref{expand1}), the damping is simply proportional to
$\alpha$.  The dependence of the growth on $\alpha$ is less clear
since $G_-^{(1)}$ itself depends on $\alpha$ in a complicated way
according to the coupled equations (\ref{expand2}) and
(\ref{expand4}).  Some insight into these equations is obtained by
considering the case $\alpha=0$, for which we find
\begin{equation}
  {{d^2W_-^{(1)}}\over{dr^2}}+
  {{d\ln({\cal I}r^3\Omega^3)}\over{dr}}{{dW_-^{(1)}}\over{dr}}+
  {{16\Sigma\Omega_{\rm b}^2}\over{{\cal I}\Omega^2}}W_-^{(1)}=
  -{{8\Omega_{\rm b}B^{(1)} W_+^{(0)*}}\over{{\cal I}r^2\Omega^3}}.
\end{equation}
For our disk model, we have approximately $\Omega\propto r^{-3/2}$,
$\Sigma\propto r^{-1/2}$, and ${\cal I}\propto r^{3/2}$ over most of
the disk.  We then obtain, approximately,
\begin{equation}
  {{d^2W_-^{(1)}}\over{dr^2}}+{{r}\over{\lambda^3}}W_-^{(1)}=
  -{{8\Omega_{\rm b}B^{(1)} W_+^{(0)*}}\over{{\cal I}r^2\Omega^3}},
  \label{airy}
\end{equation}
where
\begin{equation}
  \lambda=\left[{{\epsilon^2}\over{16(2n+3)(1+q)}}\right]^{1/3}r_{\rm b},
  \label{lambda}
\end{equation}
and $\epsilon$ is the angular semi-thickness $H/r$ of the disk.  This
is an inhomogeneous Airy equation such as is common in problems of
resonant wave excitation in differentially rotating disks.  Here, the
resonance is at the exact center of the disk, in accord with equation
(\ref{resonance}).  The forcing term on the right-hand side, however,
is proportional to $r^{5/2}$ and is therefore concentrated in the
outer parts of the disk.

\subsection{Interpretation}

The magnitude of the response $W_-^{(1)}$ (and therefore $G_-^{(1)}$)
depends on the overlap between the forcing function and the solutions
of the homogeneous equation, $Ai(-r/\lambda)$ and $Bi(-r/\lambda)$.
One may then distinguish two cases, depending on whether the outer
radius $r_2$ satisfies $r_2\gg\lambda$ or $r_2\la\lambda$.

If $r_2\gg\lambda$, the homogeneous solutions are highly oscillatory
over the disk and the overlap will be very small unless a global
resonance occurs.  This happens when there is a homogeneous solution
that (nearly) satisfies both radial boundary conditions.  This means,
in fact, that the frequency of a free bending mode of the disk (in the
inertial frame) is (nearly) equal to $2\Omega_{\rm b}$.  Then the
operator on the left-hand side of equation (\ref{airy}) is (nearly)
singular and a large response results.  This clearly occurs during the
inviscid resonances.  When this happens, equation (\ref{airy}) breaks
down; however, the analysis in Section 6 is valid.

If, instead, $r_2\la\lambda$, the response is also reasonably large
because there is little or no cancellation in the overlap integral.
This can explain the long tail of the primary resonance, where the net
growth rate is found to be positive for sufficiently small, but
non-zero, $\alpha$.

For the reference model, $\lambda\approx0.037r_{\rm b}$.  As
$r_2/r_{\rm b}$ is reduced from $0.3$ to $0.05$, we pass from the
first case $r_2\gg\lambda$, through several resonances, towards the
second case, $r_2\approx\lambda$.  The interpretation given above can
therefore explain the behavior found in Section~6.

It is natural to ask whether the parameters of real disks are likely
to allow a tilting instability in practice.  We have tested how the
the value of the outer radius at which the primary resonance occurs,
$r_2=r_{\rm p}$, varies with all the parameters of the model.  The
variations with $\epsilon$, $n$, and $q$ are well approximated by $r_{\rm
  p}\approx3.2\lambda$, $\lambda$ being given by equation
(\ref{lambda}).  The variations with $r_1/r_{\rm b}$, $w_1/r_1$,
and $w_2/r_2$ are all less significant.  Comparing this estimate
of $r_{\rm p}$ with the tidal radius $r_{\rm t}$ of a disk estimated
by Papaloizou \& Pringle (1977), we find that the inequality
\begin{equation}
  {{r_{\rm p}}\over{r_{\rm t}}}\la
  0.4\left({{\epsilon}\over{0.1}}\right)^{2/3},
  \label{stab}
\end{equation}
is satisfied for $0.2<q<10$.  We conclude that tidally truncated disks
extend too far beyond the primary resonance for instability to occur,
unless $\epsilon\ga0.4$, which is not suggested by observations
(although it should be remembered that $H$ is the true semi-thickness
of our polytropic model, and not an approximate scale-height).  For
tidally truncated disks, a tilting instability would occur only in the
unlikely case that a higher-order resonance condition were met.

It can also be seen from the above that to impose an `ingoing wave'
boundary condition at the center of the disk, as was done by
Papaloizou \& Terquem (1995), is questionable.  Those authors
envisaged that the bending wave (i.e. $W_-^{(1)}$) would be excited at
the outer edge of the disk and would propagate inwards, growing in
amplitude until nonlinear effects caused it to dissipate.  The wave
would then fail to reflect from the center of the disk.  In contrast,
we find that the wave may be considered to be launched at a local
resonance located at the center of the disk.  Since the tidal forcing
vanishes there, the wave is not significantly excited unless the width
of the resonance (proportional to $\lambda$) becomes comparable to the
radius of the disk.  In this case, the wave is launched at all radii
and global resonant effects must be taken into account.  However, the
wave amplitude does not diverge at the center of the disk; equation
(\ref{airy}) has a solution with a finite tilt and vanishing torque at
$r=0$ for the surface density profile adopted.  Nonlinear dissipation
may not occur.  Indeed, perhaps contrary to conventional wisdom, the
instability at the primary resonance (where the disk edge is at radius
$r_p$) operates in a completely inviscid disk with reflecting
boundaries.

The effects of the contribution of $W_-$ to the tilt growth can be
related back to the mode-coupling description seen in Fig.~1.  In
particular, wave equation (\ref{airy}) describes the generation of the
wave $W_-$ through the driving term on the right-hand side of that
equation.  This term involves the interaction of the $m=2$ tidal
potential, represented by $B^{(1)}$, with the rigid tilt, $W_+^{(0)}$.
This interaction produces a wave, $W_-$, of the form of an $m=1$
bending wave having frequency nearly equal to $2 \Omega_{\rm b}$ in
the inertial frame.  The interaction of the wave with the tidal field
produces a stress that corresponds to the tilt growth-rate
contribution $B^{(1)}W_-^{(1)*}$ in equation (\ref{omega2}).
Therefore the instability mechanism described by Lubow (1992) is
always at work here, but with the differences that global resonant
effects can be important, and that dissipation is not required.
Furthermore, our comparison with the damping rate induced by the $m=0$
potential indicates that the instability is suppressed for tidally
truncated disks, except in the unlikely event of a high-order
resonance.

\section{Summary and discussion}

In this paper, we have considered the linear stability of a coplanar
protostellar disk that surrounds a star in a circular-orbit binary
system.  We have determined whether a slight tilt introduced into the
disk would grow or decay in time.  The outcome depends on the size of
the disk.  For disks that are truncated by standard tidal torques,
typically resulting in an outer disk radius of about $0.3$ times the
binary separation, we find that the disk tilt generally decays in
time.

For smaller disks, tilt growth is possible.  As seen in Fig.~2, a disk
undergoes a strong, `primary' resonance with the tidal field when its
outer radius is a certain fraction of the binary separation.  This
characteristic radius, which we denote by $r_{\rm p}$, is
approximately $0.118$ times the binary separation for the parameters
we have considered (see Table~1), but would be smaller still for
thinner disks with $H/r<0.1$.  In such a resonance, the disk
experiences a growing tilt and becomes significantly warped (see
Fig.~6).  This resonance occurs when the frequency of the lowest-order
global bending mode in the disk matches the tidal forcing frequency,
which is here twice the binary orbital frequency.  Weaker resonances
occur at a series of discrete resonances corresponding to radii
greater than $r_{\rm p}$.  There is also a near resonance that occurs
close to the disk center. For disks smaller than radius $r_{\rm p}$,
this resonance causes a very slight tilt growth if $\alpha$ is
sufficiently small (but non-zero), and any initial tilt would be
retained.

For disks with radii larger than $r_{\rm p}$, including disks
truncated by standard tidal torques, the tilt will decay on
approximately the viscous time-scale of the disk, or roughly $10^3$
binary orbits for $\alpha=0.01$ (see Fig.~5).  For disks with large
tilts, nonlinear effects may shorten
the time-scale to reach small tilts, perhaps to the precessional
time-scale of the disk, or about $20$ binary orbits (Bate et~al.
2000).

The net outcome of growth or decay of the disk tilt is determined by
the competition of two torques.  As seen in the inertial frame, the
tidal torque acting on a tilted disk may be decomposed into a steady
component and an oscillatory component with twice the binary orbital
frequency.  The steady torque, resulting from the $m=0$ component of
the tidal field, causes the disk to become aligned with the binary
orbit in the presence of dissipation, while the oscillatory torque,
resulting from the $m=2$ component of the tidal field, causes
misalignment.  The steady torque produces an intuitively simple result
because it causes the disk to settle to a state of coplanarity, where
it experiences a minimum tidal potential energy, as a result of
dissipation.  The effect of the oscillatory torque is somewhat
counterintuitive, but can be understood in terms of a mode-coupling
model (see Fig.~1).  Provided that $\alpha$ is sufficiently small, the
oscillatory torque slightly dominates for smaller disks because
material in such disks is generally closer to the near resonance that
occurs in the vicinity of the disk center (see eq.~[\ref{resonance}]).

A major issue is the origin of the tilt in observed protostellar
disks.  In the case of HK Tau, the disk surrounds the secondary star,
but the two stars are similar in spectral type (Monin, M\'enard, \&
Duch\^ene 1998).  Although there are considerable uncertainties in the
system parameters, the disk could extend to its standard tidal
truncation radius, as suggested by Stapelfeldt et~al. (1998).  In that
case, the results of this paper imply that tidal effects may cause
decay of the primordial tilt, but in any case would not cause tilt
growth.  On the other hand, the existence of the tilt means that the
decay time-scale cannot be much shorter than the binary age, estimated
as $5\times10^5$~yr.  This places some constraints on both the theory
and the binary parameters, although there are considerable
uncertainties.  For example, consider the case that the binary
separation is close to its projected value of 340 AU.  For
$\alpha=0.01$, the linear tilt decay time-scale (based on Figure 2)
would be several times longer than the estimated system age. On the
other hand, the nonlinear decay time-scale estimate of Bate et al
(2000) suggests a decay time-scale substantially shorter than the
estimated age.  The nonlinear time-scale estimate would be more
compatible with a somewhat larger binary separation.

The predicted shape of a
tilted, tidally truncated disk with $H/r\approx0.1$ is not strongly
warped (see Fig.~7), in accord with the observations
(Stapelfeldt et~al. 1998; Koresko 1998).
The lack of an observed warp cannot be used as evidence
against binarity. On the other hand, a slight warp does
occur for thinner disks such as in Fig.~7 case b, which
could be observed as a small asymmetry.

Note that the decay time-scale of proper bending modes of this disk
(based on Table 3) is of order $10^4$~yr if $\alpha=0.01$, much
shorter than the linear tilt decay time-scale.  If the disk were
tilted and warped in an arbitrary way as a result of its formation
process, we would expect it to evolve rapidly to a tilted but
essentially unwarped shape, then the tilt itself would decay on a
longer time-scale.  However, the nonlinear effects discussed by Bate
et~al. (2000) are likely to speed up both stages considerably.

Similar considerations apply to a recent observational test of
coplanarity among a sample of T Tauri binaries by Donar, Jensen, \&
Mathieu (2000). The data show some evidence for approximate
coplanarity on a statistical basis. It is possible that some tidal
evolution of the tilt towards coplanarity may be have occurred, if the
tilt decays as rapidly as a disk precessional time-scale.

It is important to understand whether disk truncation could occur
close to the resonant radius $r_{\rm p}$, so that the disk would be
unstable to tilting.  For disks with a substantial tilt, Terquem
(1998) has shown that a disk of radius close to $r_{\rm p}$ is
sometimes subject to a strong resonant torque that is parallel to its
spin axis.  This resonant torque can exceed the viscous torque in the
disk for sufficiently small values of $\alpha$, $\alpha\la10^{-3}$.
If this torque could truncate an initially tilted disk at radius
$r_{\rm p}$, the disk might become strongly warped (as seen in Fig.~6)
and tilted further.  The disk radius would be less than half of the
standard tidal truncation radius.  However, it is unclear that this
torque would lead to disk truncation at $r_{\rm p}$ because it is
smoothly distributed over the disk rather than being concentrated near
$r_{\rm p}$.  This is because the resonance is global rather than
local.  The lack of a strong warp in HK Tau argues against this
process in that system.

Disks in cataclysmic binaries are expected to be much colder than
protostellar disks, having a smaller value of $H/r$.  Consequently,
such disks are even less likely to be unstable to tilting as a result
of the $m=2$ component of the tidal field (see eq.~[\ref{stab}]).

In several X-ray binaries, most notably Her X-1, there is evidence for
a tilted, precessing disk (see Wijers \& Pringle 1999 and references
therein).  The tilting mechanism we have described is very unlikely to
operate in such disks, which are expected to be tidally truncated and
to have $H/r\ll0.4$.  Therefore, it appears that tidal torques are not
responsible for the tilting of disks in X-ray binaries (cf.~Larwood
1998).  Possible mechanisms for tilting these disks include wind
torques (Schandl \& Meyer 1994) and radiation torques (Wijers \&
Pringle 1999).

Another possible application of this work is to nearly Keplerian disks
that surround black holes in active galactic nuclei. If the disk is
subject to a bar potential from the galaxy and the disk radius is
sufficiently smaller than the corotation radius of the bar, then the
disk will be subject to this tilt instability.

The results in this paper have implications to protostellar disks
perturbed by inclined planets.  A secular resonance occurs where the
precession frequency of a planet matches the local precession
frequency of an orbiting particle.  The resonant radius changes as the
nebula disperses and the resonance sweeps across a major portion of
the solar nebula (Ward 1981).  However, the current results suggest
that the effects of such resonances on the gaseous nebula are mild and
are distributed over the disk.  Further analysis can be carried out
through the methods described in this paper.

\acknowledgments

We thank Jim Pringle for encouraging this investigation and for
providing useful discussions.  We acknowledge support from NASA grant
NAG5-4310 and from the STScI visitor program.  GIO was supported by
the European Commission through the TMR network `Accretion on to Black
Holes, Compact Stars and Protostars' (contract number
ERBFMRX-CT98-0195).

\vskip2cm
\centerline{\epsfysize=10cm\epsfbox{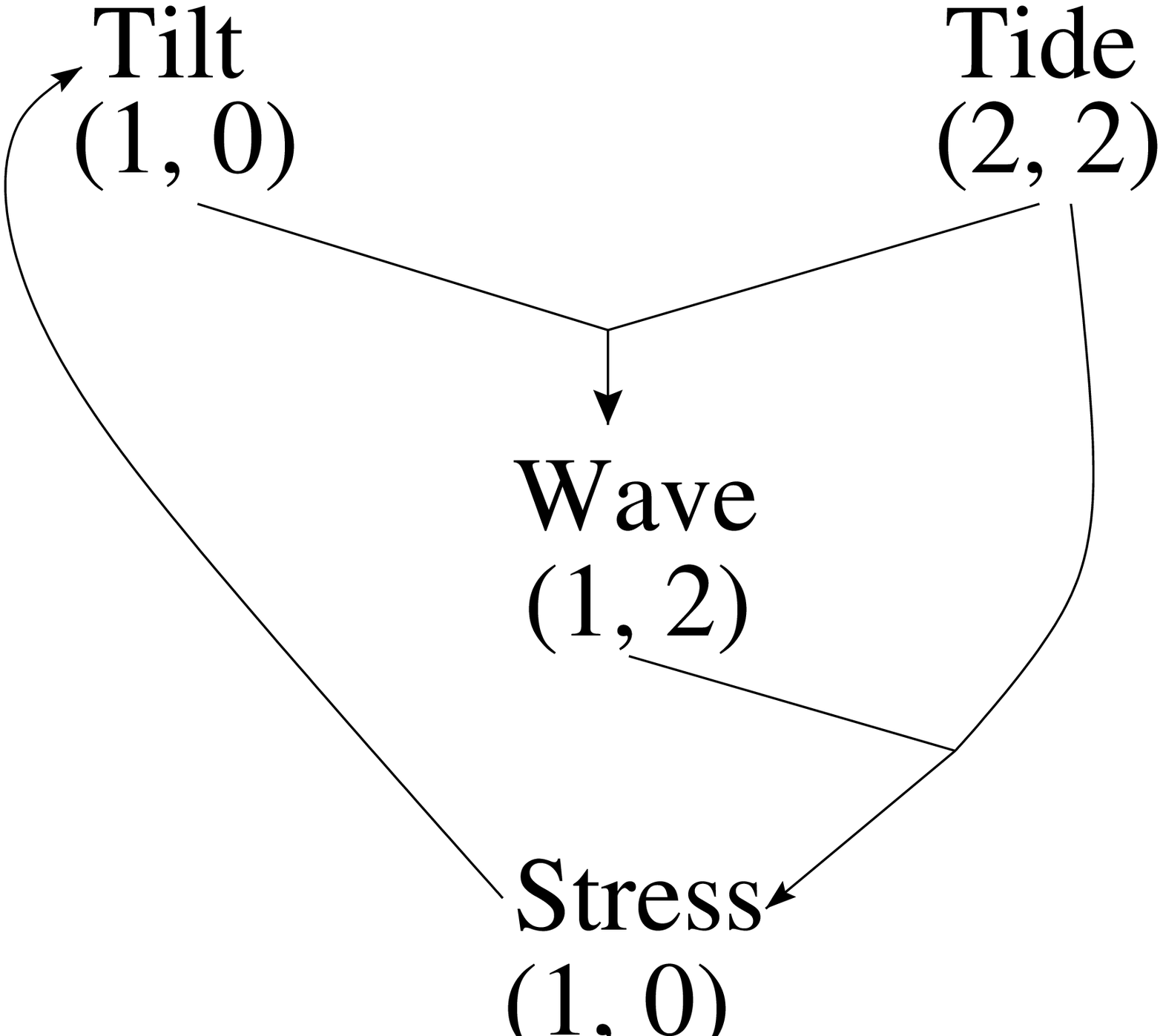}}
\vskip1cm
\figcaption[fig1.eps] {Tilt instability cycle induced by the binary
  tidal field.  We label disturbances by integer pairs $(m,l)$ for
  azimuthal and temporal dependences in the inertial frame of the form
  $\exp[i(-m\phi+l\Omega_{\rm b}t)]$, for azimuthal angle $\phi$.  The
  tilt and tide interact to produce a bending wave.  The wave and tide
  interact to produce a stress that amplifies the tilt.
  \label{fig1}}

\centerline{\epsfbox{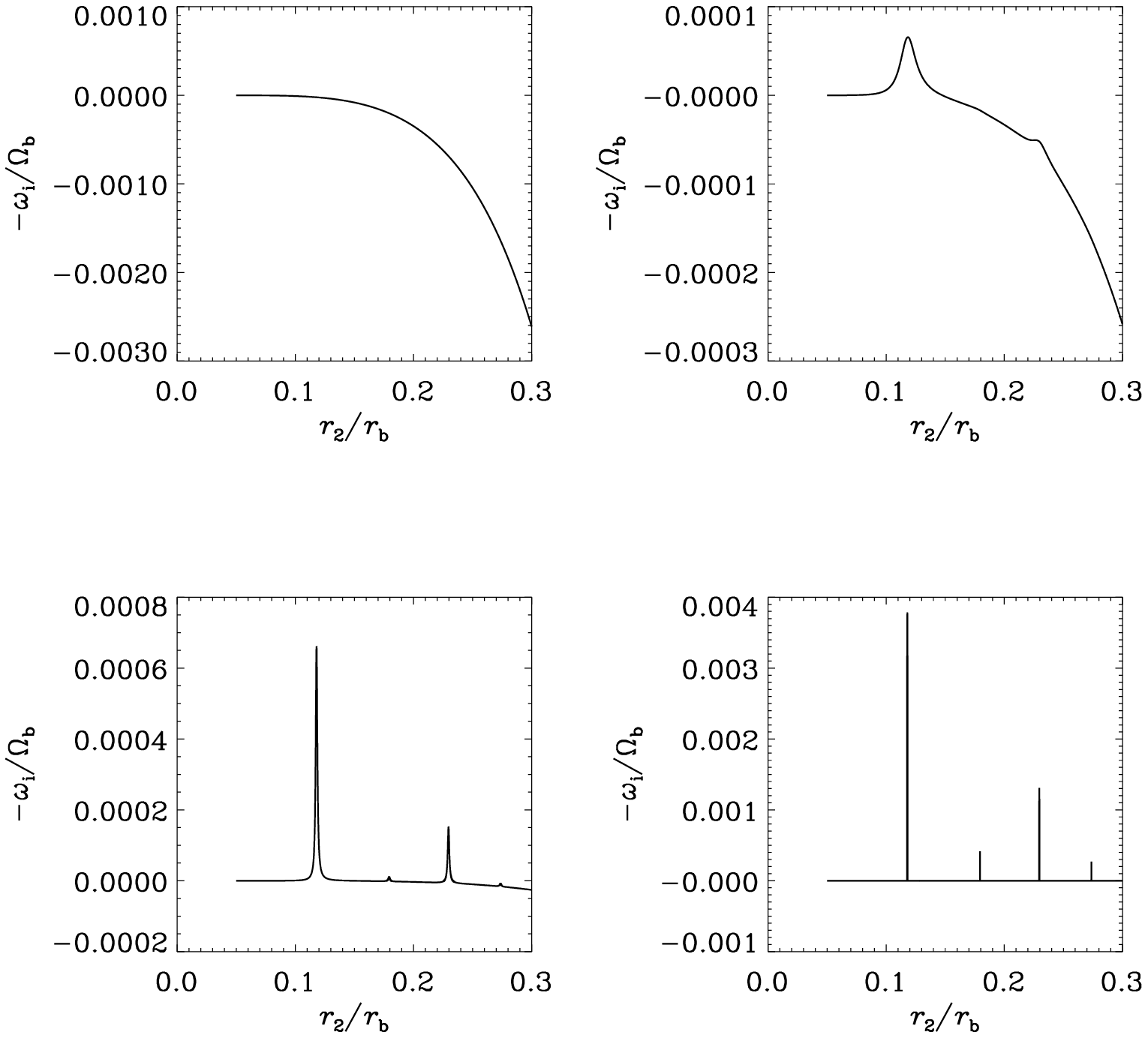}}
\figcaption[fig2.eps] {Growth rate of the modified rigid-tilt mode, in
  units of the binary frequency, plotted against the outer radius of
  the disk, in units of the binary radius.  The four panels are for
  $\alpha=0.1$ ({\it top left\/}), $\alpha=0.01$ ({\it top right\/}),
  $\alpha=0.001$ ({\it bottom left\/}), and $\alpha=0$ ({\it bottom
    right\/}); otherwise, parameters have their reference values.
  Note that the vertical scale is different in each case.
  \label{fig2}}

\centerline{\epsfbox{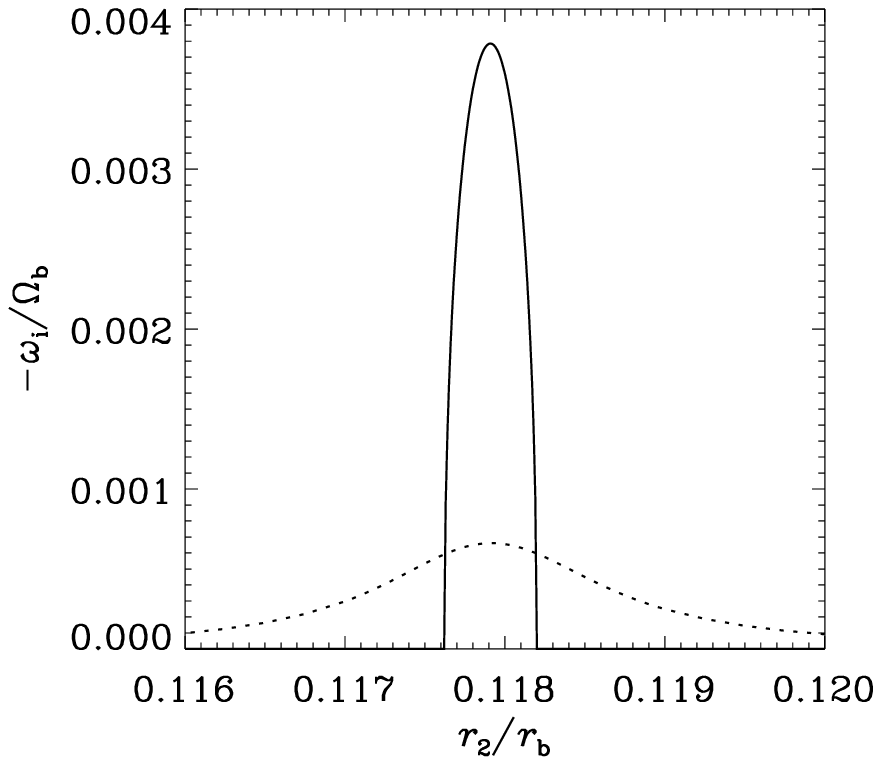}}
\figcaption[fig3.eps]
{Expanded view of Fig.~2, showing the primary resonance in the cases
  $\alpha=0$ ({\it solid line\/}) and $\alpha=0.001$ ({\it dotted
    line\/}).
  \label{fig3}}

\vskip2cm

\centerline{\epsfbox{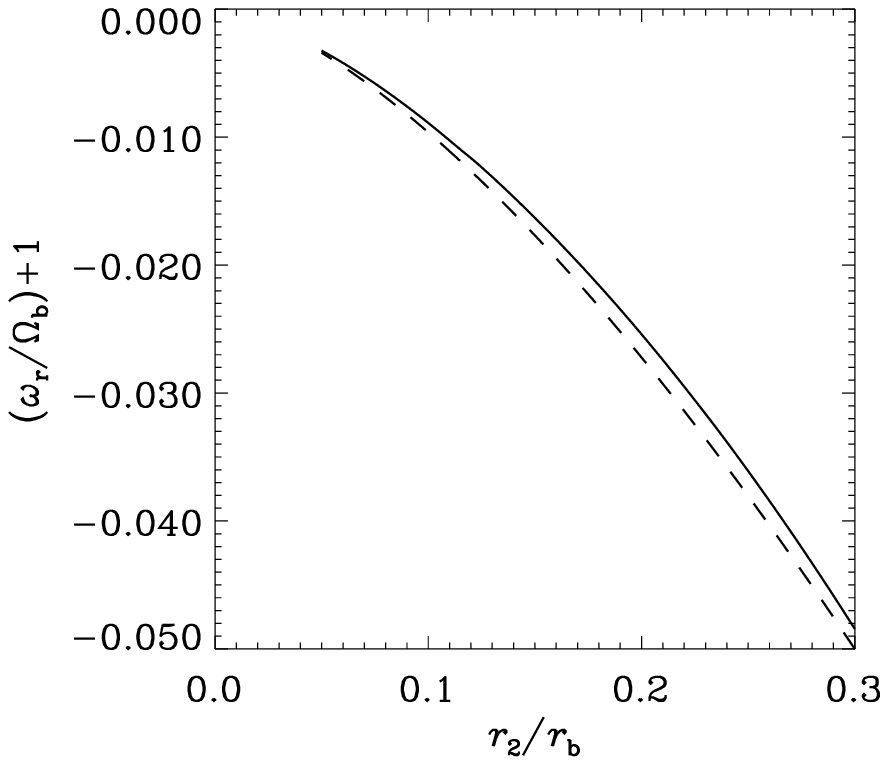}}
\figcaption[fig4.eps] {Precession frequency of the modified rigid-tilt
  mode, in units of the binary frequency, plotted against the outer
  radius of the disk, in units of the binary radius.  {\it Solid
    line\/}: reference model ($\alpha=0.01$).  {\it Dashed line\/}:
  analytic approximation from Bate et~al. (2000).
  \label{fig4}}

\centerline{\epsfbox{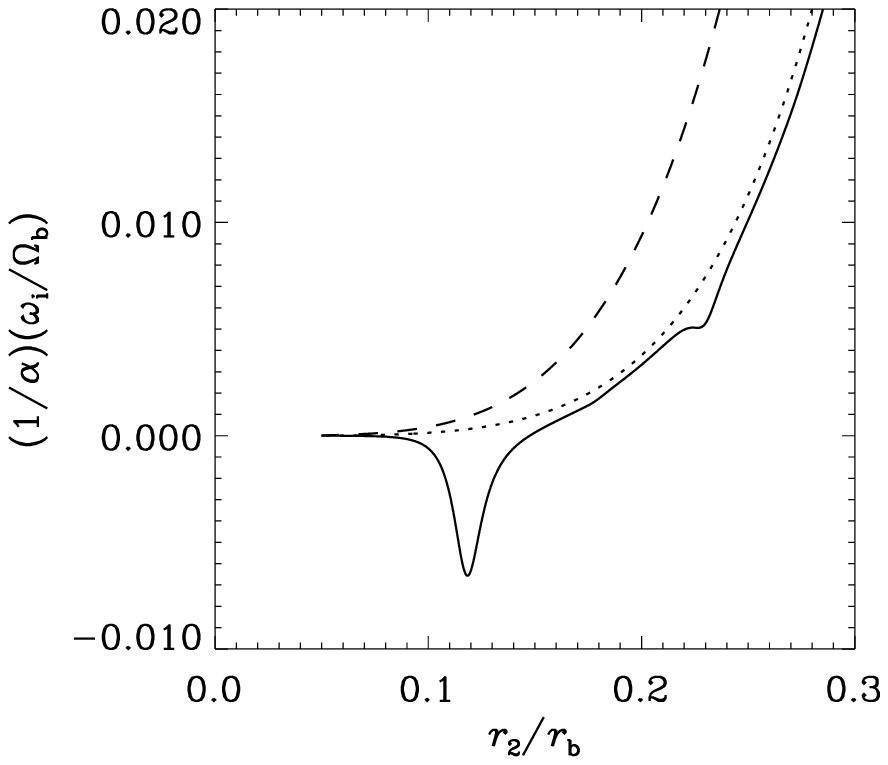}}
\figcaption[fig5.eps] {Decay rate of the modified rigid-tilt mode, in
  units of the binary frequency, divided by $\alpha$ and plotted
  against the outer radius of the disk, in units of the binary radius.
  {\it Solid line\/}: reference model ($\alpha=0.01$).  {\it Dotted
    line\/}: reference model, but with only the $m=0$ component of the
  tidal potential.  {\it Dashed line\/}: analytic approximation from
  Bate et~al.  (2000).
  \label{fig5}}

\newpage

\leftline{ }
\leftline{ }

\centerline{\epsfbox{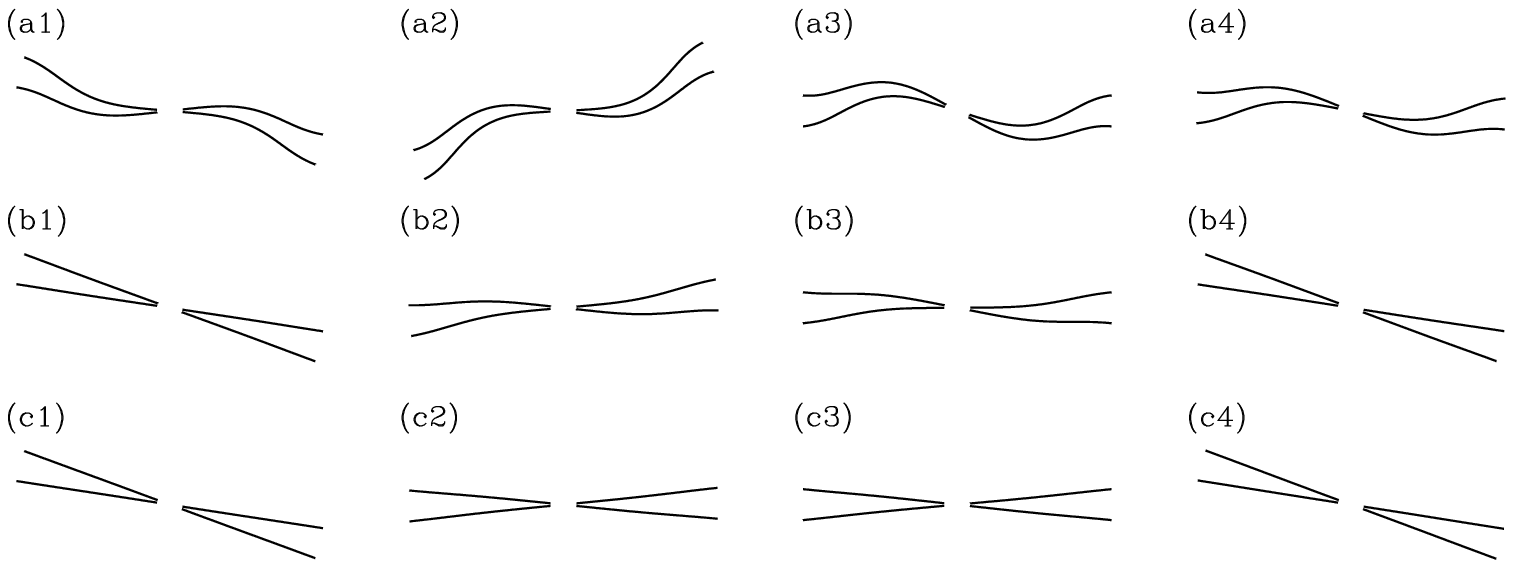}}
\figcaption[fig6.eps] {Shape of the disk while executing the modified
  rigid-tilt mode.  The disk has $r_2/r_{\rm b}=0.118$, in the middle
  of the primary resonance.  Panels (a), (b), and (c) correspond to
  the cases $\alpha=0$, $\alpha=0.001$, and $\alpha=0.01$,
  respectively, with growth rates $-\omega_{\rm i}/\Omega_{\rm
    b}=0.003596$, $0.000655$, and $0.000065$.  In each case, view (1)
  is an $xz$-section, looking along the $y$-axis, at phase $0$ of the
  cycle seen in the binary frame.  View (2) is a $yz$-section, looking
  along the negative $x$-axis, as if from the companion star.  Panels
  (3) and (4) are the same as (1) and (2), but at phase $\pi/2$ in the
  cycle.  The eigenfunctions have been renormalized for ease of
  comparison.
  \label{fig6}}

\vskip2cm

\centerline{\epsfbox{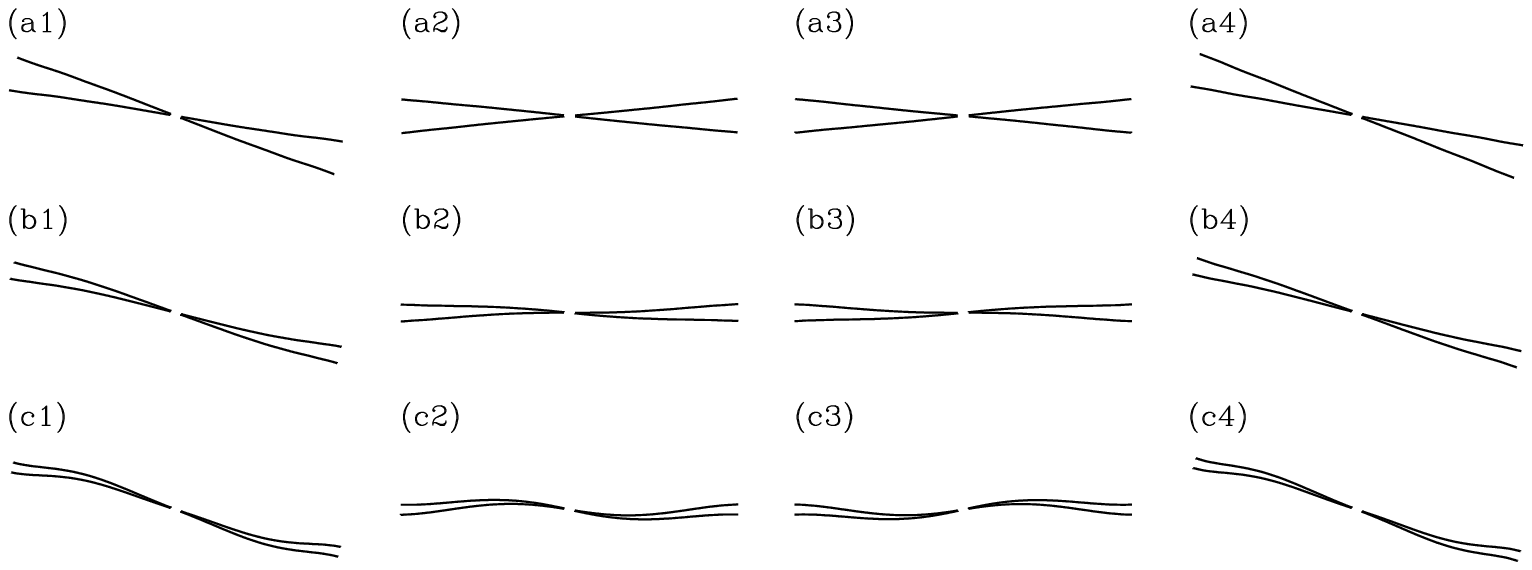}}
\figcaption[fig7.eps] {As for Fig.~6, but for a disk with $r_2/r_{\rm
    b}=0.3$.  Here $\alpha=0.01$ is fixed and panels (a), (b), and (c)
  correspond to the cases $\epsilon=0.1$, $\epsilon=0.05$, and
  $\epsilon=0.03$, respectively, with damping rates $\omega_{\rm
    i}/\Omega_{\rm b}=0.000258$, $0.001205$, and $0.003601$.
  \label{fig7}}

\newpage

\appendix

\section{Derivation of the reduced equations for linear bending waves}

Let the small parameter $\epsilon$ be a characteristic value of the
angular semi-thickness $H/r$ of the disk.  Then set
\begin{eqnarray}
  \kappa^2&=&\Omega^2\left[1+\epsilon f_\kappa(r)\right],\\
  \Omega_z^2&=&\Omega^2\left[1+\epsilon f_z(r)\right],\\
  \alpha&=&\epsilon f_\alpha(r),\\
\end{eqnarray}
where the functions $f$ are $O(1)$, which includes the possibility of
their being arbitrarily small.  The disk is assumed to satisfy the
Navier-Stokes equation with a (dynamic) shear viscosity given by
\begin{equation}
  \mu={{\alpha p}\over{\Omega}}.
\end{equation}

To describe the internal structure of the disk, adopt units in which
the radius of the disk and the orbital frequency are $O(1)$.
Introduce the stretched vertical coordinate
\begin{equation}
  \zeta={{z}\over{\epsilon}},
\end{equation}
which is $O(1)$ inside the disk.  We then find, for the unperturbed disk,
\begin{eqnarray}
  u&=&O(\epsilon^3),\\
  v&=&r\Omega(r)+\epsilon^2r\Omega_2(r,\zeta)+O(\epsilon^3),\\
  w&=&O(\epsilon^4),\\
  \rho&=&\epsilon^s\left[\rho_0(r,\zeta)+\epsilon\rho_1(r,\zeta)+
  O(\epsilon^2)\right],\\
  p&=&\epsilon^{s+2}\left[p_0(r,\zeta)+\epsilon p_1(r,\zeta)+
  O(\epsilon^2)\right],\\
  \mu&=&\epsilon^{s+3}\left[\mu_0(r,\zeta)+O(\epsilon)\right],
\end{eqnarray}
where $s$ is an arbitrary positive parameter.  Any viscous evolution
of the disk occurs on a long time-scale $O(\epsilon^{-3})$ and is
consistently neglected.  The vertical component of the equation of
motion implies, at $O(\epsilon)$,
\begin{equation}
  0=-{{1}\over{\rho_0}}{{\partial p_0}\over{\partial\zeta}}-\Omega^2\zeta,
\end{equation}
and, at $O(\epsilon^2)$,
\begin{equation}
  0=-{{1}\over{\rho_0}}{{\partial p_1}\over{\partial\zeta}}+
  {{\rho_1}\over{\rho_0^2}}{{\partial p_0}\over{\partial\zeta}}-
  f_z\Omega^2\zeta,
\end{equation}
while the radial component at $O(\epsilon^2)$ gives
\begin{equation}
  -2r\Omega\Omega_2=-{{1}\over{\rho_0}}{{\partial p_0}\over{\partial r}}-
  \Omega{{d\Omega}\over{dr}}\zeta^2.
\end{equation}

Consider linear bending waves with azimuthal wavenumber $m=1$ in which
the Eulerian perturbation of $u$, say, is
\begin{equation}
  {\rm Re}\left[u^\prime(r,z,t)\,e^{-i\phi}\right].
\end{equation}
It is known that these waves travel radially at a speed comparable to
the sound speed (Papaloizou \& Lin 1995).  Therefore the
characteristic time-scale for the evolution of the warped shape is the
radial sound crossing time $r/c_{\rm s}\approx\epsilon^{-1}\Omega^{-1}$,
implying that the perturbations evolve on a time-scale
$O(\epsilon^{-1})$ that is long compared to the orbital time-scale
[$O(1)$] but much shorter than the viscous time-scale
[$O(\epsilon^{-3})$].  This is captured by a slow time coordinate
\begin{equation}
  T=\epsilon t.
\end{equation}
For the perturbations, introduce the scalings
\begin{eqnarray}
  u^\prime&=&\epsilon u_1^\prime(r,\zeta,T)+
  \epsilon^2 u_2^\prime(r,\zeta,T)+O(\epsilon^3),\\
  v^\prime&=&\epsilon v_1^\prime(r,\zeta,T)+
  \epsilon^2 v_2^\prime(r,\zeta,T)+O(\epsilon^3),\\
  w^\prime&=&\epsilon w_1^\prime(r,\zeta,T)+
  \epsilon^2 w_2^\prime(r,\zeta,T)+O(\epsilon^3),\\
  \rho^\prime&=&\epsilon^s\left[\rho_1^\prime(r,\zeta,T)+
  \epsilon\rho_2^\prime(r,\zeta,T)+O(\epsilon^2)\right],\\
  p^\prime&=&\epsilon^{s+2}\left[p_1^\prime(r,\zeta,T)+
  \epsilon p_2^\prime(r,\zeta,T)+O(\epsilon^2)\right].
\end{eqnarray}
The overall amplitude of the perturbations is of course arbitrary
since this is a linear analysis.  \footnote{The scaling adopted here
  is, however, of some significance, since it corresponds to a
  (differential) tilt angle comparable to the angular thickness of the
  disk.  This is appropriate for a warp of observational consequence.
  We note that nonlinear effects may be significant for warps of this
  amplitude.  Unfortunately, the Eulerian perturbation method used
  here tends to overestimate the degree of nonlinearity.  For example,
  although the fractional Eulerian density perturbation is of order
  unity, the dominant perturbation is (locally) a rigid tilt and the
  Lagrangian density perturbation is in fact of higher order in
  $\epsilon$.  Nonlinear effects can occur, however, owing to the fact
  that the horizontal motions are comparable to the sound speed.  In
  particular, these motions can be damped by a parametric instability
  (Gammie, Goodman, \& Ogilvie 2000; Bate et~al. 2000).  The Eulerian
  method remains the most convenient way of obtaining the equations if
  one is satisfied with a formal linearization.  Since we are
  considering a stability problem in this paper, this method is
  sufficient for our purposes.}

The perturbed equations for $w$, $\rho$, and $p$ at leading order are
\begin{eqnarray}
  -i\Omega w_1^\prime+{{1}\over{\rho_0}}
  {{\partial p_1^\prime}\over{\partial\zeta}}-
  {{\rho_1^\prime}\over{\rho_0^2}}
  {{\partial p_0}\over{\partial\zeta}}&=&0,\\
  -i\Omega\rho_1^\prime+w_1^\prime{{\partial\rho_0}\over{\partial\zeta}}+
  \rho_0{{\partial w_1^\prime}\over{\partial\zeta}}&=&0,\\
  -i\Omega p_1^\prime+w_1^\prime{{\partial p_0}\over{\partial\zeta}}+
  \gamma p_0{{\partial w_1^\prime}\over{\partial\zeta}}&=&0,
\end{eqnarray}
where $\gamma$ is the adiabatic exponent.  These may be combined to
give
\begin{equation}
  {{\partial}\over{\partial\zeta}}
  \left(\gamma p_0{{\partial w_1^\prime}\over{\partial\zeta}}\right)=0.
\end{equation}
The general solution, regular at the disk surface, is
\begin{eqnarray}
  w_1^\prime&=&ir\Omega W,\\
  \rho_1^\prime&=&r{{\partial\rho_0}\over{\partial\zeta}}W,\\
  p_1^\prime&=&r{{\partial p_0}\over{\partial\zeta}}W,
\end{eqnarray}
where $W(r,T)$ is a dimensionless complex function to be determined.
These perturbations correspond to applying a rigid tilt to each
annulus of the disk.  The tilt varies with radius and time according
to the function $W(r,T)$, which is related to the unit tilt vector
$\bell$ through
\begin{equation}
  W=\ell_x+i\ell_y.
\end{equation}

The perturbed equations for $u$ and $v$ at leading order are
\begin{eqnarray}
  -i\Omega u_1^\prime-2\Omega v_1^\prime&=&0,\\
  -i\Omega v_1^\prime+{\textstyle{{1}\over{2}}}\Omega u_1^\prime&=&0.
\end{eqnarray}
The general solution is
\begin{eqnarray}
  u_1^\prime&=&U,\\
  v_1^\prime&=&-{\textstyle{{1}\over{2}}}iU,
\end{eqnarray}
where $U(r,\zeta,T)$ is a complex function to be determined.

The perturbed equations for $w$, $\rho$, and $p$ at the next order are
\begin{eqnarray}
  -i\Omega w_2^\prime+{{1}\over{\rho_0}}
  {{\partial p_2^\prime}\over{\partial\zeta}}-
  {{\rho_2^\prime}\over{\rho_0^2}}
  {{\partial p_0}\over{\partial\zeta}}&=&F_w,
  \label{fw}\\
  -i\Omega\rho_2^\prime+
  w_2^\prime{{\partial\rho_0}\over{\partial\zeta}}+
  \rho_0{{\partial w_2^\prime}\over{\partial\zeta}}&=&F_\rho,\\
   -i\Omega p_2^\prime+
  w_2^\prime{{\partial p_0}\over{\partial\zeta}}+
  \gamma p_0{{\partial w_2^\prime}\over{\partial\zeta}}&=&F_p,
  \label{fp}
\end{eqnarray}
where
\begin{eqnarray}
  F_w&=&-ir\Omega{{\partial W}\over{\partial T}}-
  {{\rho_1}\over{\rho_0}}r\Omega^2W-
  {{1}\over{\rho_0}}{{\partial\rho_0}\over{\partial\zeta}}
  f_zr\Omega^2\zeta W,\\
  F_\rho&=&-r{{\partial\rho_0}\over{\partial\zeta}}
  {{\partial W}\over{\partial T}}-
  ir\Omega{{\partial\rho_1}\over{\partial\zeta}}W-
  {{1}\over{r}}{{\partial}\over{\partial r}}(\rho_0rU)+
  {{\rho_0U}\over{2r}},
\end{eqnarray}
and $F_p$ will not be required.  Now the linear operator defined by
the left-hand sides of equations (\ref{fw})--(\ref{fp}) is singular
owing to the existence of the tilt mode identified above.  The
corresponding solvability condition is
\begin{equation}
  \int\left(i\rho_0F_w+F_\rho\Omega\zeta\right)\,d\zeta=0,
\end{equation}
where the integral is over the entire vertical extent of the disk.
This evaluates to
\begin{equation}
  \Sigma_0r\Omega\left(2{{\partial W}\over{\partial T}}+
  if_z\Omega W\right)-
  {{1}\over{r^2}}{{\partial}\over{\partial r}}
  \int\rho_0r^2\Omega U\zeta\,d\zeta=0,
\end{equation}
where
\begin{equation}
  \Sigma_0=\int\rho_0\,d\zeta
\end{equation}
is the surface density.

The perturbed equations for $u$ and $v$ at the next order are
\begin{eqnarray}
  -i\Omega u_2^\prime-2\Omega v_2^\prime&=&F_u,\\
  -i\Omega v_2^\prime+{\textstyle{{1}\over{2}}}\Omega u_2^\prime&=&F_v,
\end{eqnarray}
where
\begin{eqnarray}
  F_u&=&-{{\partial U}\over{\partial T}}-
  {{1}\over{\rho_0}}{{\partial}\over{\partial r}}
  \left(r{{\partial p_0}\over{\partial\zeta}}W\right)+
  {{r}\over{\rho_0^2}}{{\partial\rho_0}\over{\partial\zeta}}
  {{\partial p_0}\over{\partial r}}W+
  {{1}\over{\rho_0}}{{\partial}\over{\partial\zeta}}
  \left(\mu_0{{\partial U}\over{\partial\zeta}}\right),\\
  F_v&=&{\textstyle{{1}\over{2}}}i{{\partial U}\over{\partial T}}-
  {\textstyle{{1}\over{2}}}f_\kappa\Omega U-
  ir^2\Omega{{\partial\Omega_2}\over{\partial\zeta}}W+
  {{i}\over{\rho_0}}{{\partial p_0}\over{\partial\zeta}}W-
  {{i}\over{2\rho_0}}{{\partial}\over{\partial\zeta}}
  \left(\mu_0{{\partial U}\over{\partial\zeta}}\right).
\end{eqnarray}
Again the linear operator is singular, with the solvability condition
\begin{equation}
  F_u+2iF_v=0.
\end{equation}
This evaluates to
\begin{equation}
  -2{{\partial U}\over{\partial T}}-if_\kappa\Omega U+
  {{2f_\alpha}\over{\rho_0\Omega}}{{\partial}\over{\partial\zeta}}
  \left(p_0{{\partial U}\over{\partial\zeta}}\right)+
  r\Omega^2\zeta{{\partial W}\over{\partial r}}=0.
\end{equation}
By inspection, the solution is of the form $U\propto\zeta$, with
\begin{equation}
  \left(2{{\partial}\over{\partial T}}+
  if_\kappa\Omega+2f_\alpha\Omega\right)
  \left({{U}\over{\zeta}}\right)=
  r\Omega^2{{\partial W}\over{\partial r}}.
\end{equation}
If we now define
\begin{equation}
  G={{{\cal I}_0r^2\Omega}\over{2}}\left({{U}\over{\zeta}}\right),
\end{equation}
where
\begin{equation}
  {\cal I}_0=\int\rho_0\zeta^2\,d\zeta
\end{equation}
is the second vertical moment of the density, we obtain the coupled
equations
\begin{equation}
  \Sigma_0r^2\Omega\left({{\partial W}\over{\partial T}}+
  {\textstyle{{1}\over{2}}}if_z\Omega W\right)=
  {{1}\over{r}}{{\partial G}\over{\partial r}},
\end{equation}
\begin{equation}
  {{\partial G}\over{\partial T}}+
  {\textstyle{{1}\over{2}}}if_\kappa\Omega G+f_\alpha\Omega G=
  {{{\cal I}_0r^3\Omega^3}\over{4}}{{\partial W}\over{\partial r}}.
\end{equation}

Finally, if we step back from the asymptotic analysis and present the
equations in physical terms, we obtain
\begin{equation}
  \Sigma r^2\Omega\left[{{\partial W}\over{\partial t}}+
  \left({{\Omega_z^2-\Omega^2}\over{\Omega^2}}\right)
  {{i\Omega}\over{2}}W\right]=
  {{1}\over{r}}{{\partial G}\over{\partial r}},
\end{equation}
\begin{equation}
  {{\partial G}\over{\partial t}}+
  \left({{\kappa^2-\Omega^2}\over{\Omega^2}}\right)
  {{i\Omega}\over{2}}G+\alpha\Omega G=
  {{{\cal I}r^3\Omega^3}\over{4}}{{\partial W}\over{\partial r}},
\end{equation}
where now
\begin{eqnarray}
  \Sigma&=&\int\rho\,dz,\\
  {\cal I}&=&\int\rho z^2\,dz.
\end{eqnarray}


\begin{references}
  \reference{BBCLOPT00} Bate, M.~R., Bonnell, I.~A., Clarke, C.~J.,
    Lubow, S.~H., Ogilvie, G.~I., Pringle, J.~E., \& Tout, C.~A. 2000,
    \mnras, submitted
  \reference{BGT84} Borderies, N., Goldreich, P., \& Tremaine, S. 1984,
    \apj, 284, 429
  \reference{B96} Burrows, C.~J. et al. 1996, \apj, 473, 437
  \reference{DI97} Demianski, M., \& Ivanov, P. 1997, \aap, 324, 829
  \reference{DJM} Donar, A., Jensen, E.~L.~N. Jensen, \& Mathieu, R.~D. 2000, 
    BAAS, in press
  \reference{GGO00} Gammie, C. F., Goodman, J., \& Ogilvie, G. I. 2000,
    \mnras, submitted
  \reference{GNM93} Ghez, A.~M., Neugebauer, G., \& Matthews, K. 1993,
    \aj, 106, 2005
  \reference{GT81} Goldreich, P., \& Tremaine, S. 1981, \apj, 243, 1062
  \reference{JMF96} Jensen, E.~L.~N., Mathieu, R.~D., \& Fuller, G.~A. 1996, 
    \apj, 458, 312
  \reference{KAGM82} Katz, J.~I., Anderson, S.~F., Grandi, S.~A., \&
    Margon, B. 1982, \apj, 260, 780
  \reference{K98} Koresko, C.~D. 1998, \apj, 507, L145
  \reference{L98} Larwood, J.~D. 1998, \mnras, 299, L32
  \reference{LNPT96} Larwood, J.~D., Nelson, R.~P., Papaloizou,
    J.~C.~B., \& Terquem, C. 1996, \mnras, 282, 597
  \reference{L91} Lubow, S.~H. 1991, \apj, 381, 259
  \reference{L92} Lubow, S.~H. 1992, \apj, 398, 525
  \reference{MO96} McCaughrean, M.~J., \& O'Dell, C.~R. 1996, \aj,
    111, 1977
  \reference{M94} Mathieu, R.~D. 1994, \araa, 32, 465
  \reference{MMD98} Monin, J.-L., M\'enard, F., \& Duch\^ene, G. 1998,
    \aap, 339, 113
  \reference{MA98} Murray, J.~R., \& Armitage, P.~J. 1998, \mnras,
    300, 561
  \reference{O99} Ogilvie, G.~I. 1999, \mnras, 304, 557
  \reference{O00} Ogilvie, G.~I. 2000, \mnras, submitted
  \reference{O94} Osaki, Y. 1996, \pasp, 108, 39
  \reference{OB95} Osterloh, M., \& Beckwith, S.~V.~W. 1995, \apj,
    439, 288
  \reference{P77} Paczy\'nski, B. 1977, \apj, 216, 822
  \reference{PL95} Papaloizou, J.~C.~B., \& Lin, D.~N.~C. 1995, \apj,
    438, 841
  \reference{PP77} Papaloizou, J.~C.~B., \& Pringle, J.~E. 1977, \mnras,
    181, 441
  \reference{PP83} Papaloizou, J.~C.~B., \& Pringle, J.~E. 1983, \mnras,
    202, 1181
  \reference{PT95} Papaloizou, J.~C.~B., \& Terquem, C. 1995, \mnras,
    274, 987
  \reference{P92} Pringle, J.~E. 1992, \mnras, 258, 811
  \reference{SM94} Schandl, S., Meyer, F. 1994, \aap, 289, 149
  \reference{SKMBPB98} Stapelfeldt, K.~R., Krist, J.~E., M\'enard, F.,
    Bouvier, J., Padgett, D.~L., \& Burrows, C.~J. 1998, \apj, 502, L65
  \reference{T99} Terquem, C.~E.~J.~M.~L.~J. 1998, \apj, 509, 819
  \reference{WP99} Wijers, R. A. M. J., Pringle, J. E. 1999, \mnras,
    308, 207
\end{references}
\end{document}